%% file: main.tex
\documentclass[letterpaper,12pt,peerreviewca,draftcls]{IEEEtran}
\usepackage{csm16}
\usepackage[margin=1in]{geometry}
\usepackage[nolists,nomarkers,tablesfirst]{endfloat} 
\usepackage{amsmath} 

\usepackage{url}
\usepackage{graphicx}
\usepackage[table,xcdraw]{xcolor}
\usepackage{longtable}
\usepackage{verbatim}
\makeatletter
\newcommand{\verbatimfont}[1]{\def\verbatim@font{#1}}%
\makeatother
\verbatimfont{\ttfamily\small}

\newcommand{\bi}{\begin{itemize}}\newcommand{\ei}{\end{itemize}}
\newcommand{\be}{\begin{equation}}\newcommand{\ee}{\end{equation}}
\newcommand{\bee}{\begin{enumerate}}\newcommand{\eee}{\end{enumerate}}
\newcommand{\bea}{\begin{eqnarray}}\newcommand{\eea}{\end{eqnarray}}
\newcommand{\beas}{\begin{eqnarray*}}\newcommand{\eeas}{\end{eqnarray*}}
\newcommand{\bc}{\begin{center}}\newcommand{\ec}{\end{center}}
\usepackage[left,pagewise]{lineno}
\usepackage[english]{babel}
\usepackage{blindtext}
\usepackage{multirow}
\usepackage{array}
\usepackage{tabularx}
\usepackage{tikz}
\usetikzlibrary{positioning,shapes,arrows}

\title{A Realistic Simulation Testbed of A Turbocharged Spark-Ignited Engine System\\
\Large A platform for the evaluation of fault diagnosis algorithms and strategies}
\author{Kok Yew Ng, Erik Frisk, Mattias Krysander, and Lars Eriksson\\
	POC: K.\ Y.\ Ng (mark.ng@ulster.ac.uk)\\ \today }

\newif\ifPDF \ifx\pdfoutput\undefined\PDFfalse \else\ifnum\pdfoutput > 0\PDFtrue \else\PDFfalse \fi \fi
\ifPDF
\usepackage[pdftex, plainpages = false, colorlinks=true, linkcolor=black, citecolor = green!50!blue, urlcolor = blue, filecolor=black, pagebackref=false, hypertexnames=false,  pdfpagelabels ]{hyperref}
\fi

\begin{document}

\pagestyle{empty}
\noindent \copyright ~2020 IEEE.  Personal use of this material is permitted.  Permission from IEEE must be obtained for all other uses, in any current or future media, including reprinting/republishing this material for advertising or promotional purposes, creating new collective works, for resale or redistribution to servers or lists, or reuse of any copyrighted component of this work in other works. \\~\\

\noindent This is a peer-reviewed and accepted version of the following in press document. \\~\\

\noindent {\large K. Y. Ng, E. Frisk, M. Krysander, and L. Eriksson. ``A Realistic Simulation Testbed of A Turbocharged Spark-Ignited Engine System: A Platform for the Evaluation of Fault Diagnosis Algorithms and Strategies,'' {\it IEEE Control Systems (In Press)}, vol. 40, no. 2, 2020.}
\newpage

\maketitle
\CSMsetup

The study of fault diagnosis on automotive engine systems has been an interesting and ongoing topic for many years. Numerous research was conducted by both the automakers and research institutions to discover new and more advanced methods to perform diagnosis for better fault isolation (FI). Some of the research reported in this field has been reported in \cite{gerler1995model,kher2001automobile,nyberg2002model,murphey2003automotive,denton2016advanced}.

In most automotive systems today, the diagnostic systems monitor multiple components in the engine and are   independent of each other. However, some faults have a tendency to manifest and trigger several other monitors either simultaneously or subsequently \cite{goodloe2010monitoring}. For instance, a disconnected intake system hose has a high potential to result in both flow and pressure faults further along in the engine system. To overcome this, residuals from several monitors (coupled with an intelligent algorithm) are needed to enhance the accuracy in isolating the faults, for both locations and identifying the root cause of the problem. The ability to identify the root cause of the fault and pinpoint its exact location to the correct component is crucial for taking proper measures and avoiding the replacement of misdiagnosed engine components \cite{scacchioli2006model}.

One of the main disadvantages of existing diagnostic systems is that the faults are not detected in a chronological order. As a result (depending on the locations of the monitors in the engine and the propagation time of electrical signals), the manifested fault(s) may trigger the monitors much sooner than the root cause of the problem. Also, should a monitor break down or not run well, it may take more time to detect the root fault, or worse, the fault may not be detected at all. This will in turn lead to incorrect onboard diagnostic reconfiguration efforts or an incorrect replacement of the so-called ”faulty” components offboard by the technician \cite{weber1999multiple,da2012knowledge}. These misdiagnosis and robustness issues are especially critical in autonomous vehicular systems, where it is essential for the computers onboard the vehicles to know the health of the system, such that corrective measures can be taken to protect the lives of occupants as well as other road users. Depending on the severity of the fault, the vehicle can either be reconfigured to operate at a reduced performance level (to ensure safety until the vehicle is brought into the workshop for repair and maintenance works) or safely brought to a halt at a suitable location as soon as possible. Reports on similar concepts of reconfigurations for the purpose of fault-tolerant control, self-healing, and recoverability of autonomous systems can be found in \cite{tang2008prognostics,loureiro2012bond,Loureiro2014integration}. The failure to detect and to isolate a fault (or incorrectly identify one) may cause the reconfiguration of the system to not be optimized. This affects the health and lifespan of other components or the engine as a whole. Therefore, an improved fault diagnosis method is crucial to not only identify the root cause of the problem, but also to immediately and correctly reconfigure the engine before the condition worsens. While many existing software and simulation packages provide interesting simulated studies on the dynamics of the engine system, very few have explored the design and analysis of fault diagnosis schemes. See \cite{isermann1999hardware,butler1999matlab,assanis2000validation,lee2003cost,yoon2005development}.

This article presents a simulation testbed of the engine system, whereby its operation can be realistically simulated using industrial-standard driving cycles such as the Worldwide harmonized Light vehicles Test Procedures (WLTP), New European Driving Cycle (NEDC), Extra-Urban Driving Cycle (EUDC), and EPA Federal Test Procedure (FTP-75) \cite{kuhlwein2014development}. A GUI interface enables the user to set simulation preferences such as the desired driving cycle as well as one of the 11 faults of interest. The performance of the developed fault diagnosis scheme can be analyzed without physically inducing the faults on the physical engine, thus minimizing the risk of shortening its lifespan or causing permanent damages to the engine. The simulation environment (available as opensource at \url{https://github.com/nkymark/TCSISimTestbed}), is intended to enhance the development of theoretical and applications of fault diagnosis of engine systems, with hope that researchers in the field encourage research collaborations or use it as a virtual laboratory for teaching purposes.

\section{Modeling The Engine}\label{EngineDes}
\begin{figure}[t!]
\begin{center}
\includegraphics[width=\textwidth]{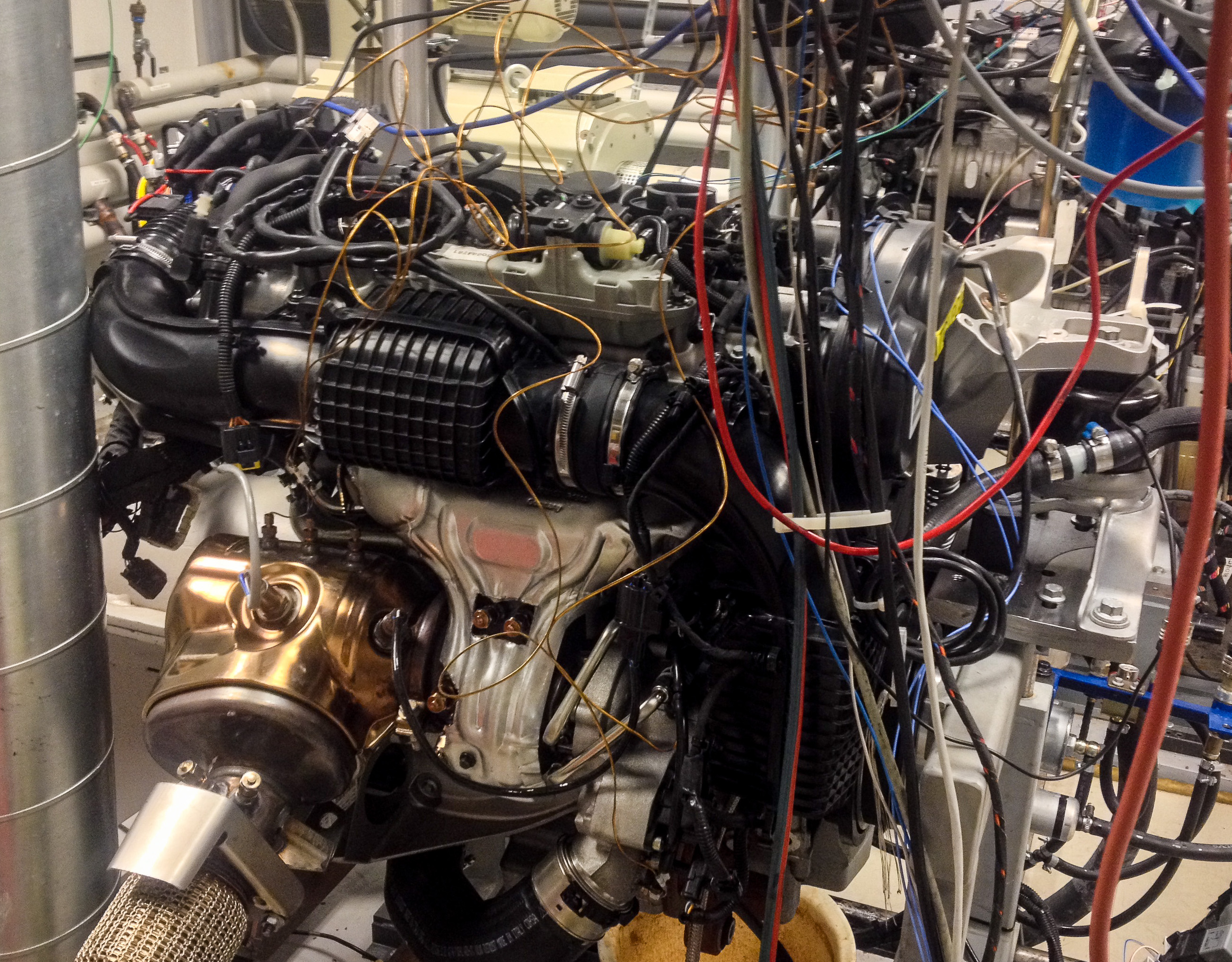}
\caption{\label{Bench}The actual engine test bench used for data collection in the lab.}
\end{center}
\end{figure}

\begin{figure}[t!]
\begin{center}
\fbox{\includegraphics[width=\textwidth]{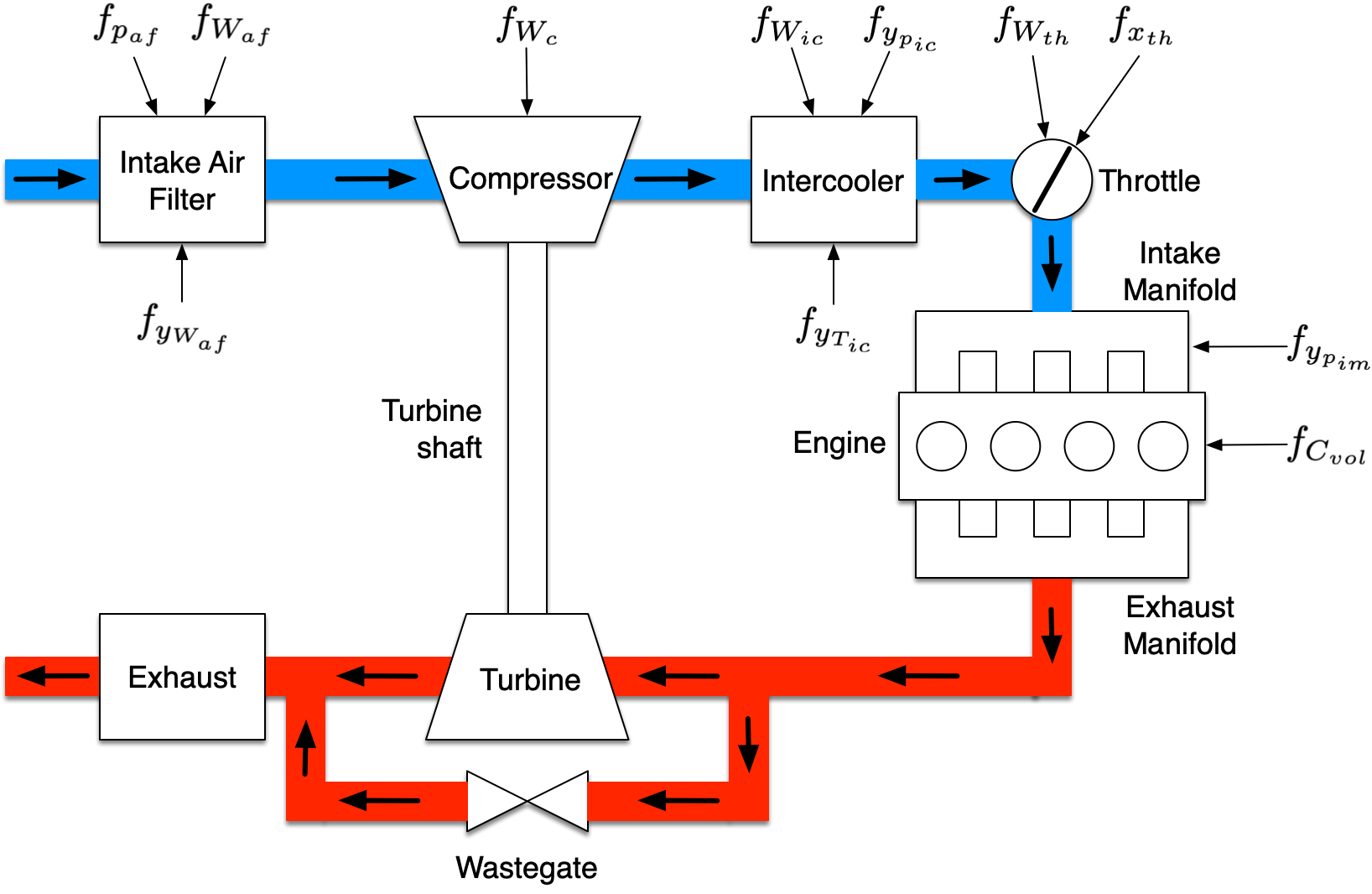}}
\caption{\label{EngineSch}Schematic diagram of the engine system with its subsystems. The blue paths indicate the mass flow of air before entering the engine block, while the red paths show the usually much hotter gases produced as a result of the combustion reactions in the engine block. The figure also shows the locations in the engine system where each fault is induced.}
\end{center}
\end{figure}

The simulation environment uses a four-cylinder single turbocharged spark-ignited (TCSI) petrol engine as the testbed to design and verify the performance of fault diagnosis schemes. Figure \ref{Bench} shows the actual engine test bench used for data collection in the lab, while Figure \ref{EngineSch} shows the schematic diagram of the engine system, which consists of the following subsystems:
\begin{itemize}
	\item Air filter: Ambient air enters the engine system, and the filter prevents abrasive particulate matters from entering the engine block.
	\item Compressor: Modeled based on the radial compressor and driven by the turbine, air from the air filter is compressed, thus increasing the air volumetric flow, pressure, and temperature.
	\item Intercooler: Air from the compressor is cooled down while maintaining the air mass flow velocity.
	\item Throttle: Used to control the pressure in the intake manifold, thus regulating the amount of fuel that goes into the engine.
	\item Intake manifold: The combustion mixture of air and fuel is distributed evenly to the four cylinders in the engine.
	\item Engine block: The combustion mixture is ignited to generate the torque for mechanical work.
	\item Exhaust manifold: Directs the gases produced by the combustion reactions in the engine to the turbine and the wastegate.
	\item Turbine: Harvests the energy of gases from the exhaust manifold to generate power to drive the compressor.
	\item Wastegate: A valve that bypasses the turbine and controls the power delivered by the turbocharger.
	\item Exhaust system: Gases in the engine system exits to the ambient environment.
\end{itemize}

The engine system is modeled using differential equations that describe the air flow through the subsystems in the engine. These equations are derived based on the mean value engine model (MVEM) for a TCSI engine, as reported in \cite{Lars, EriNie:2014}. The key parameters of the vehicle and the engine used for this testbed (as well as the total model equations of the engine with a list of engine variables) can be found in ``\nameref{AppA}'' and ``\nameref{AppB}.'' The system has 13 states: $\{T_{af}, p_{af}, T_{c}, p_{c}, T_{ic}, p_{ic}, T_{im}, p_{im}, T_{em}, p_{em}, T_{t}, p_{t}, \omega_{t}\}$, which are the temperatures (K) and pressures (Pa) in the air filter, compressor, intercooler, intake manifold, exhaust manifold, and turbine, as well as the turbine speed (rad/s), respectively. There are six actuators: $\{A_{th}, u_{wg}, \omega_{eREF}, \lambda, p_{amb}, T_{amb}\}$, which represent the throttle position area (m$^{2}$), wastegate input ([$0...1$]), reference engine speed (rad/s), air-fuel ratio, ambient pressure (Pa), and temperature (K), respectively. The system has nine sensors: $\{T_{c}, p_{c}, T_{ic}, p_{ic}, T_{im}, p_{im}, p_{em}, W_{af}, Tq_{e}\}$, where $W_{af}$ and $Tq_{e}$ are the mass flow in the air filter (kg/s) and the engine torque (N$\cdot$m), respectively (see Table \ref{ActSen}).

\section{Generation of Reference Inputs and Controller Design}
\label{Controller}
\begin{figure}[t!]
\begin{center}
\fbox{\input{./Figures/Figure3}}
\caption{\label{Engine}The closed-loop engine control system with the boost controller, actuators, and sensors.}
\end{center}
\end{figure}
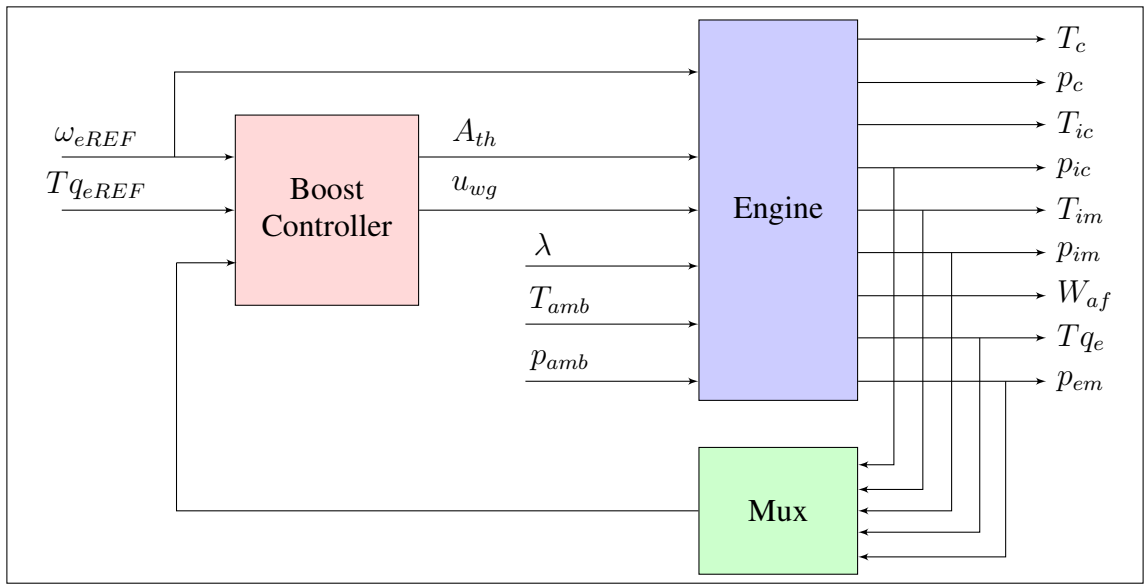

This section discusses the design of the proportional-integral (PI)-based boost controller with anti-windup to generate the control inputs for the throttle effective area $A_{th}$ and the wastegate actuator for the turbocharger $u_{wg}$. Figure \ref{Engine} shows the closed-loop engine control system with the boost controller. See Tables \ref{EngineVar} and \ref{EnginePara} in ``\nameref{AppB}'' for the descriptions and values of the engine variables and parameters used in the following equations.

To estimate the gear shift points, it is assumed that the speed of the moving vehicle per 1000 rpm in 8th gear is 62.9 km/h. From the key vehicle parameters in Table \ref{VehiclePara}, the vehicle speed per 1000 rpm (km/h), $v_{g, 1000rpm}$ for each gear is
\begin{equation}
v_{g, 1000rpm} = \frac{120 \pi r_w}{\mbox{final gear ratio} \times \mbox{current gear ratio}},
\end{equation}
where $r_{w}$ is the wheel radius (m).

The results are then tabulated in Table \ref{GearShifts}, which also shows the estimated gear shift points of the gearbox. The data in Tables \ref{VehiclePara} and \ref{GearShifts} (together with the information of vehicle speed versus time from the driving cycle profile) provide the reference engine speed (rad/s), $\omega_{eREF}$ and reference engine torque (N$\cdot$m), $Tq_{eREF}$ for the boost controller. The reference engine speed is
\begin{equation}
\omega_{eREF} = \frac{V i_{gear}(V)}{r_w},
\end{equation}
where $V$ is the vehicle speed (m/s) obtained from the driving cycle profile and the function $i_{gear}(V)$ is the gear shifting vector developed from $V$ and the gear ratios in Table \ref{VehiclePara}.

To obtain the reference engine torque, $Tq_{eREF}$, the force equation of the vehicle is first expressed using
\begin{equation}
m_v \dot V = F_w - F_d - F_r,
\end{equation}
where $F_w, F_d,$ and $F_r$ are the forces (N) at the wheel, drag resistance force, and roll resistance force, respectively, and $m_{v}$ is the mass of the vehicle (kg). The forces $F_d$ and $F_r$ can then be further determined using
\begin{align}
F_d &= \frac{1}{2} \rho_a c_d A_f V^2, \\
F_r &= m_v c_r g,
\end{align}
where $\rho_a = 1.29$ kg/m$^3$ is the air density, $g$ is gravity (m/s$^{2}$), $c_{d}$ the drag coefficient, and $A_{f}$ represents the frontal area of the vehicle (m$^{2}$). If the torque produced at the wheel is written as $Tq_w = F_w r_w,$
then the reference engine torque can be finally expressed as
\begin{equation}
Tq_{eREF} = \frac{Tq_w}{i_{gear}(V)}.
\end{equation}

To model the driver accelerator pedal interpretation, the reference brake mean effective pressure (BMEP) can first be expressed using
\begin{equation}
BMEP_{REF} = \frac{2\pi n_{r} Tq_{eREF}}{V_{d}} \label{eq:BMEPref},
\end{equation}
where $Tq_{eREF}$ is the reference engine torque (N$\cdot$m), $V_{d}$ the displacement volume of the engine (m$^{3}$), and $n_{r}$ the number of revolutions per power stroke of the engine (for a four-cyclinder engine, $n_{r} = 2$). As a result, the reference intake manifold and intercooler pressures ($p_{imREF}$ and $p_{icREF}$, respectively) are obtained as
\begin{eqnarray}
p_{imREF} &=& \frac{BMEP_{REF} + C_{P0}}{C_{P1}}, \\
p_{icREF} &=& p_{imREF} + \Delta p_{thREF},
\end{eqnarray}
where $\Delta p_{thREF}$ is the regulated pressure drop across the throttle (Pa). The constants $C_{P0}$ and $C_{P1}$ are computed as
\begin{equation}
\left[\begin{array}{c} C_{P0} \\ C_{P1} \end{array}\right] = \frac{2\pi n_{r} Tq_{e}}{V_{d}} \left[\begin{array}{cc} -1 & p_{im}\end{array}\right]^{+},
\end{equation}
where $Tq_{e}$ is the measured engine torque (N$\cdot$m) and $p_{im}$ is the measured intake manifold pressure (Pa).

The reference throttle effective area (m$^{2}$) $A_{thREF}$ is then computed as
\begin{equation}
A_{thREF} = W_{eiREF} \frac{\sqrt{R_{a}T_{amb}}}{\Psi_{thREF}}, \label{eq:AthREF}
\end{equation}
where $R_{a}$ is the gas constant (J/(kg$\cdot$K)), $T_{amb}$ the ambient temperature, and $\Psi_{thREF}$ is the reference throttle flow coefficient (\%).	The reference mass flow into the engine (kg/s) $W_{eiREF}$ and $\Psi_{thREF}$ are computed as
\begin{eqnarray}
W_{eiREF} &=& \frac{C_{\eta_{vol}}V_{d}~\omega_{eREF}p_{imREF}}{4\pi R_{a}(r_{c} - 1)T_{im}} \left( r_{c} - \left(\frac{p_{em}}{p_{imREF}} \right)^{\kappa_{em}} \right), \\
\Psi_{thREF} &=& \Pi_{thREF} \sqrt{\frac{2\kappa_{th}}{\kappa_{th}-1}\left( \Pi_{thREF}^{\frac{2}{\kappa_{th}}}  - \Pi_{thREF}^{\frac{\kappa_{th}+1}{\kappa_{th}}}\right)},
\end{eqnarray}
where $C_{\eta_{vol}}$ is the volumetric efficiency constant, $r_{c}$ the compression ratio, $T_{im}$ the intake manifold temperature (K), $p_{em}$ the exhaust manifold pressure (Pa), $\kappa_{em}$ the ratio of specific heats at the exhaust, and $\kappa_{th}$ the ratio of specific heats at the throttle. The pressure ratio in the throttle $\Pi_{thREF}$ is obtained as
\begin{equation}
\Pi_{thREF} = \frac{p_{imREF}}{max(p_{imREF},p_{ic})},
\end{equation}
where $p_{ic}$ is the intercooler pressure (Pa).

To design the controller with anti-windup for the throttle, the reference throttle position $\alpha_{thREF}$ is computed as
\begin{equation}
\alpha_{thREF} = \alpha_{thFF} + \alpha_{thFB},
\end{equation}
where $\alpha_{thFF}$ and $\alpha_{thFB}$ are the feedforward and feedback components of the controller for the throttle position, respectively. Using the solution from (\ref{eq:AthREF}), $\alpha_{thFF}$ is expressed as
\begin{equation}
\alpha_{thFF} = - \frac{a_{0}}{2a_{2}} \pm \sqrt{\frac{A_{thREF}-a_{0}}{a_{2}}+\left(\frac{a_{1}}{a_{2}}\right)^{2}},
\end{equation}
where the constants $a_{0}, a_{1}$, and $a_{2}$ are parameters obtained from measurements in the engine lab. The feedforward component of the controller enables it to respond quickly to changes made to the engine (for example, a rapid acceleration when the accelerator pedal is depressed fully onto the floor).

The feedback component of the controller $\alpha_{thFB}$ is obtained as
\begin{equation}
\alpha_{thFB} = K_{p,th}e_{im} + \frac{K_{p,th}}{T_{i,th}}\int{e_{im} + K_{p,th}(\alpha_{thREF\_SAT} - \alpha_{thREF})~}{dt} ,
\end{equation}
where $K_{p,th}$ and $T_{i,th}$ are the proportional and integral gains of the feedback controller, respectively, and $e_{im} = p_{imREF} - p_{im}$. The saturation of the reference throttle position $\alpha_{thREF\_SAT}$ is defined as the static nonlinearity
\begin{equation}
\alpha_{thREF\_SAT} = \left\{ \begin{array}{ll} \alpha_{thMAX}, & \mbox{if~~} \alpha_{thREF} > \alpha_{thMAX} \\ \alpha_{thREF}, & \mbox{if~~} \alpha_{thMIN} < \alpha_{thREF} < \alpha_{thMAX} \\ \alpha_{thMIN}, & \mbox{if~~} \alpha_{thREF} < \alpha_{thMIN} \end{array} \right. ,
\end{equation}
where $\alpha_{thMAX}$ and $\alpha_{thMIN}$ are the maximum and minimum allowed actuation signals for the throttle position, respectively. The feedback component of the controller ensures that the engine system is able to follow its references during operation.

The controller for the wastegate input consists of only a feedback component, and it is expressed using
\begin{equation}
u_{wgFB} = K_{p,wg}e_{ic} + \frac{K_{p,wg}}{T_{i,wg}}\int{e_{ic} + K_{p,wg}(u_{wgREF\_SAT} - u_{wgREF})~}{dt} ,
\end{equation}
where $K_{p,wg}$ and $T_{i,wg}$ are the proportional and integral gains of the feedback controller, respectively, and $e_{ic} = p_{ic} - p_{icREF}$. The saturation of the reference wastegate input $u_{wgREF\_SAT}$ is defined as the static nonlinearity
\begin{equation}
u_{wgREF\_SAT} = \left\{ \begin{array}{ll} u_{wgMAX}, & \mbox{if~~} u_{wgREF} > u_{wgMAX} \\ u_{wgREF}, & \mbox{if~~} u_{wgMIN} < u_{wgREF} < u_{wgMAX} \\ u_{wgMIN}, & \mbox{if~~} u_{wgREF} < u_{wgMIN} \end{array} \right. ,
\end{equation}

\begin{figure}[t!]
\begin{center}
\includegraphics[width=\textwidth]{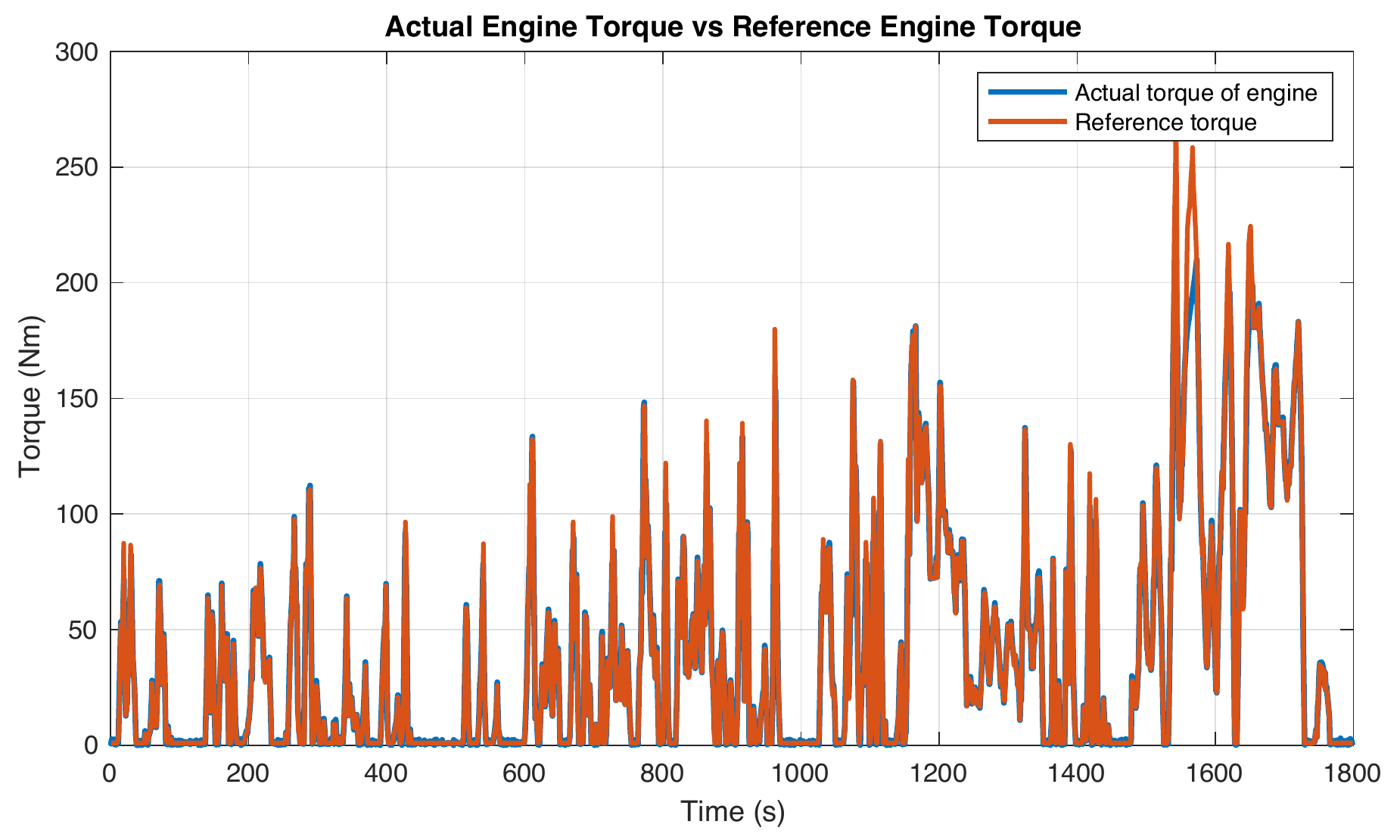}
\caption{\label{Ref}Simulation result showing the engine torque from the model follows its reference during the Worldwide harmonized Light vehicles Test Procedures (WLTP) driving cycle profile.}
\end{center}
\end{figure}

The design of the controller for the engine is then verified in simulations using the reference engine torque $Tq_{eREF}$ and speed $\omega_{eREF}$, generated from a selected driving cycle. Figure \ref{Ref} shows the reference and actual torque of the engine during the WLTP driving cycle. It can be seen that the actual engine torque is able to follow its reference well.

\begin{table}[t!]
\caption{\label{ActSen}The system states, actuators, and sensor measurements of the engine system in Figure \ref{EngineSch}.}
\centering
\begin{tabularx}{0.7\textwidth}{Xl} \hline
Description & Unit \\ \hline
System states: & \\
~~~Air filter temperature, $T_{af}$						& K 			\\
~~~Air filter pressure, $p_{af}$						& Pa			\\
~~~Compressor temperature, ${T_{c}}$					& K 			\\
~~~Compressor pressure, ${p_{c}}$ 						& Pa 			\\
~~~Intercooler temperature, ${T_{ic}}$					& K 			\\
~~~Intercooler pressure, ${p_{ic}}$ 					& Pa 			\\
~~~Intake manifold temperature, ${T_{im}}$ 				& K 			\\
~~~Intake manifold pressure, ${p_{im}}$ 				& Pa 			\\
~~~Exhaust manifold temperature, ${T_{em}}$ 			& K 			\\
~~~Exhaust manifold pressure, ${p_{em}}$ 				& Pa 			\\
~~~Turbine temperature, ${T_{t}}$ 						& K 			\\
~~~Turbine pressure, ${p_{t}}$ 							& Pa 			\\
~~~Turbine speed, $\omega_{t}$							& rad/s			\\
 & \\
Actuators: & \\
~~~Reference engine speed, $\omega_{eREF}$				& rad/s 		\\
~~~Control input for throttle position area, $A_{th}$ 	& m$^2$ 		\\
~~~Control input for wastegate, $u_{wg}$ 				& $[0...1]$ 	\\
~~~Air-fuel ratio, $\lambda$ 							& $[-]$ 		\\
~~~Ambient pressure, $p_{amb}$  						& Pa 			\\
~~~Ambient temperature, $T_{amb}$ 						& K 			\\
 &  \\
Sensors: & \\
~~~Compressor temperature, ${T_{c}}$					& K 			\\
~~~Compressor pressure, ${p_{c}}$ 						& Pa 			\\
~~~Intercooler temperature, ${T_{ic}}$					& K 			\\
~~~Intercooler pressure, ${p_{ic}}$ 					& Pa 			\\
~~~Intake manifold temperature, ${T_{im}}$ 				& K 			\\
~~~Intake manifold pressure, ${p_{im}}$ 				& Pa 			\\
~~~Air filter mass flow, ${W_{af}}$ 					& kg/s 			\\
~~~Engine torque, ${Tq_{e}}$ 							& N$\cdot$m 	\\
~~~Exhaust manifold pressure, ${p_{em}}$ 				& Pa 			\\ \hline
\end{tabularx}
\end{table}

\begin{figure}[t!]
\begin{center}
\includegraphics[width=\textwidth]{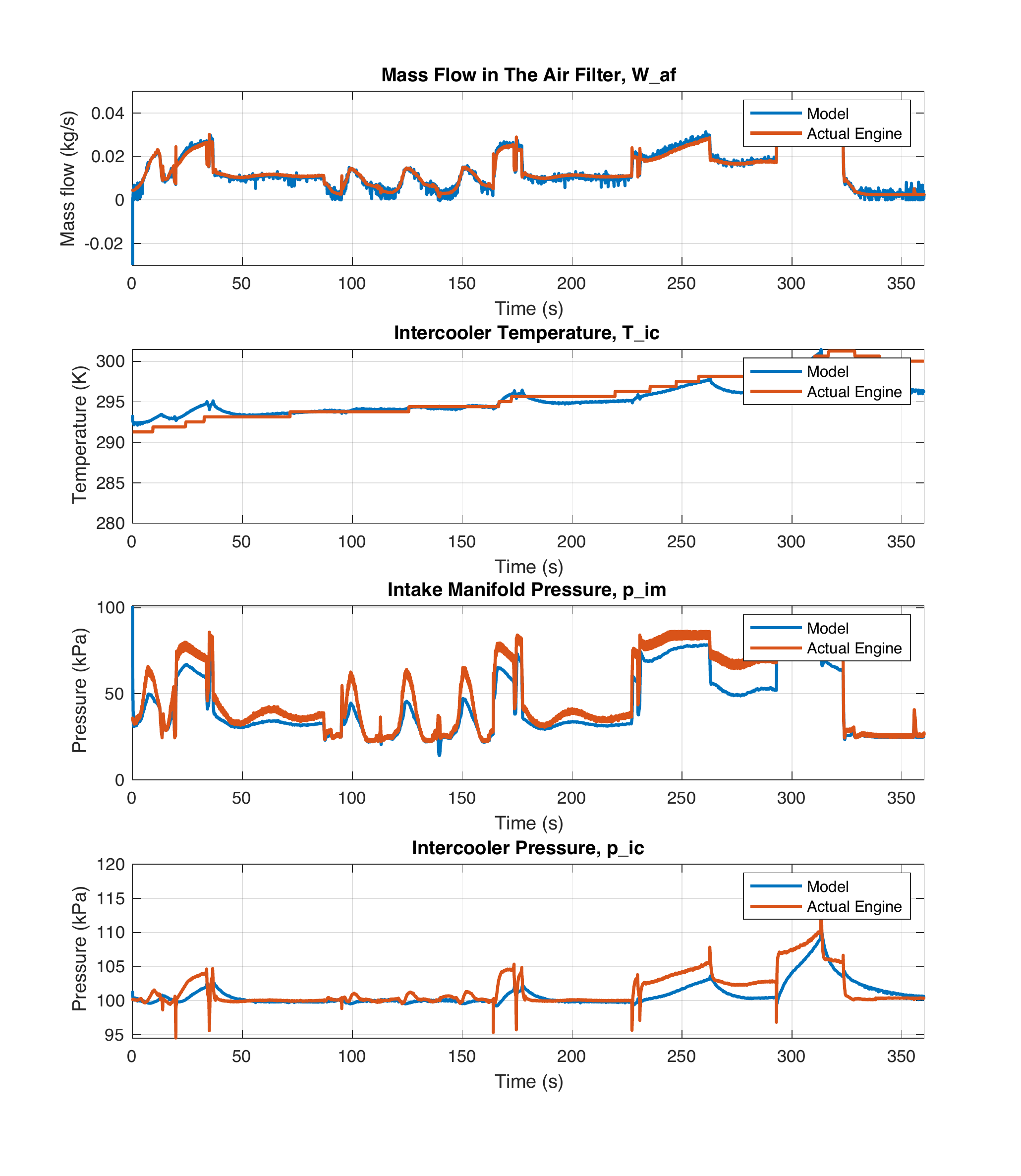}
\caption{\label{EUDC}Plots showing sensor signals of the model (blue lines) against the actual sensors of the engine in the lab (red lines) during the Extra-Urban Driving Cycle (EUDC) driving cycle profile. The sensor measurements are (from top to bottom) $y_{W_{af}}, y_{T_{ic}}, y_{p_{im}}$, and $y_{p_{ic}}$. The figures show that the engine model is realistic and accurately represents the actual engine system.}
\end{center}
\end{figure}

The engine model is also verified against the actual test bench system to ensure that the model is realistic and viable for simulations of real-world operations. Considering that the engine is a system with 13 states and is highly nonlinear (with many interconnected subsystems where air-flow can travel upstream and downstream depending on the pressure difference), it has always been a challenge to accurately model an entire engine system. The dynamics of the engine model are compared with those of the actual test bench system in the lab, which are fully controlled in real time and have sensor measurements visualized and recorded using a dSPACE MicroAutoBox + RapidPro system as well as INCA tools. The sensor measurements recorded for comparison between the model and the test bench are the air filter mass flow $y_{W_{af}}$, intercooler temperature $y_{T_{ic}}$, intake manifold pressure $y_{p_{im}}$, and intercooler pressure $y_{p_{ic}}$. During the EUDC run, the engine model produces sensor measurements that are close to the actual test bench system. These results show that the engine model is realistic and accurately represents the actual engine system (see Figure \ref{EUDC}).

\section{Fault Scenarios}\label{Faults}
The simulation testbed considers 11 sensor, actuator, and variable faults of different degrees of severity in different parts of the engine system: six variable faults ($f_{p_{af}}, f_{C_{vol}}, f_{W_{af}}, f_{W_{th}}, f_{W_{c}}, f_{W_{ic}}$), one actuator measurement fault ($f_{x_{th}}$), and four sensor measurement faults ($f_{y_{W_{af}}}, f_{y_{p_{im}}},f_{y_{p_{ic}}},f_{y_{T_{ic}}}$). Some faults are less severe, and the engine can be reconfigured to a reduced performance operation mode to accommodate the faults until the vehicle is sent for repair and maintenance. Some other faults are more severe that, if not detected and isolated promptly, might cause permanent and serious damages to the engine system (which in turn will endanger the occupants in the vehicle as well as other road users).

\begin{table}[t!]
\caption{\label{tab:fault}Faults, their descriptions, and types.}
\begin{tabularx}{\textwidth}{lXXXl} \hline
Fault				& Description & Fault Threshold & Nature of Fault (Active Period) & Severity \\ \hline
$f_{p_{af}}$		& Loss of pressure in the air filter 	 					& 20 kPa pressure drop	 			& Abrupt (From 200 s till $T_{DC}$) 			& Medium \vspace{3mm} \\
$f_{C_{vol}}$ 		& Intake valve timing stuck at arbitrary position 			& Stuck at end or middle positions 	& Abrupt pulses (Active for 30 s every 150 s)  	& High \vspace{3mm} \\
$f_{W_{af}}$		& Air leakage between the air filter and the compressor 	& 20\% of flow through leakage 		& Incipient (From 200 s till $T_{DC}$) 			& Medium \vspace{3mm} \\
$f_{W_{c}}$			& Air leakage between the compressor and the intercooler 	& 20\% of flow through leakage 		& Abrupt (From 0.4$T_{DC}$ till $T_{DC}$)		& High \vspace{3mm} \\
$f_{W_{ic}}$		& Air leakage between the intercooler and the throttle 		& 20\% of flow through leakage 		& Abrupt (From 0.4$T_{DC}$ till 0.8$T_{DC}$) 	& High \vspace{3mm} \\
$f_{W_{th}}$ 		& Air leakage after the throttle in the intake manifold		& 20\% of flow through leakage 		& Abrupt pulses (Active for 40 s every 200 s) 	& High \vspace{3mm} \\
$f_{x_{th}}$ 		& Throttle position actuator error							& Fault leading to 20\% flow error 	& Abrupt (From 0.4$T_{DC}$ till $T_{DC}$) 		& Medium \vspace{3mm} \\
$f_{y_{W_{af}}}$ 	& Air filter flow sensor fault								& 20\% flow error 					& Abrupt pulses (Active for 30 s every 150 s)	& Low \vspace{3mm} \\
$f_{y_{p_{im}}}$	& Intake manifold pressure sensor fault 					& 20\% pressure deviation 			& Incipient (From 200 s till $T_{DC}$)			& Low \vspace{3mm} \\
$f_{y_{p_{ic}}}$ 	& Intercooler pressure sensor fault 						& 20\% pressure deviation 			& Abrupt pulses (Active for 40 s every 200 s)	& Low \vspace{3mm} \\
$f_{y_{T_{ic}}}$ 	& Intercooler temperature sensor fault 						& 20 K offset 						& Abrupt pulses (Active for 30 s every 150 s) 	& Low \\ \hline
\end{tabularx}
\begin{flushleft}
*$T_{DC}$ represents the duration of the driving cycle: WLTP (1800 s), NEDC (1220 s), EUDC (400 s), FTP-75 (1874 s).
\end{flushleft}
\end{table}

\subsection{Fault Types and Classification}
The faults included in this simulation testbed can be categorized into three types: sensor fault, actuator fault, and variable fault.

\subsubsection{Sensor Faults}
This research considers four sensor measurement faults: $f_{y_{W_{af}}}, f_{y_{p_{im}}},f_{y_{p_{ic}}}$, and $f_{y_{T_{ic}}}$. The nature of these faults is due to electrical or mechanical errors that lead to either an offset or a deviation in the sensor measurements.

The $f_{y_{W_{af}}}$ fault indicates a sensor measurement error in the air filter flow. The air filter flow sensor measures the amount of air that goes into the engine. As such, it is critical for this fault to be fixed and the necessary parts replaced as soon as it is detected.

The remaining sensor faults are pressure and temperature measurement errors in the intercooler ($f_{y_{p_{ic}}}, f_{y_{T_{ic}}}$) and intake manifold ($f_{y_{p_{im}}}$) of the engine. The pressure measurement errors ($f_{y_{p_{ic}}}$ and $f_{y_{p_{im}}}$) produce a deviation of 20\% in the measured values. The $f_{y_{p_{im}}}$ is modeled using a long-term incipient fault to indicate a drift in the sensor signal over time, while the $f_{y_{p_{ic}}}$ is modeled using repeating abrupt pulses. The $f_{y_{T_{ic}}}$ indicates an offset in the sensor that measures the temperature in the intercooler. This fault is also modeled using repeating abrupt pulses.

\subsubsection{Actuator Faults}
The actuator fault considered in this research is the $f_{x_{th}}$, which indicates a throttle position actuator error, where an angular fault in the actuator leads to a flow error. This fault is modeled using repeating abrupt pulses. As this fault directly affects the throttle (and subsequently, the amount of fuel used for combustion), its severity level is medium.

\subsubsection{Variable Faults}
This research considers six variable faults: $f_{p_{af}}, f_{C_{vol}}, f_{W_{af}}, f_{W_{th}}, f_{W_{c}}$, and $f_{W_{ic}}$. The nature of these faults is due to physical and mechanical damages to certain parts of the engine system, leading to either a pressure drop, leakages, or a performance degradation of the engine itself. As a result, most of these faults are of high severity.

The $f_{p_{af}}$ fault indicates a pressure drop in the air filter due to a restriction in the flow. The fault is modeled using a long-term abrupt fault to simulate a constant restrictive flow in the air filter.

The fault $f_{C_{vol}}$ indicates that the intake valve timing actuator is stuck at an arbitrary position. As a result, this affects the volumetric efficiency of the engine, which subsequently affects the overall performance of the engine system such as power output, emission control, and fuel consumption. The volumetric efficiency is modeled as a function of the intake valve timing actuator position. Therefore, this is a serious fault and must be quickly detected and isolated. This fault is modeled using repeating abrupt pulses.

The remaining variable faults are leakages that could happen in different parts of the engine system, such as the air filter ($f_{W_{af}}$), compressor ($f_{W_{c}}$), intercooler ($f_{W_{ic}}$), and throttle ($f_{W_{th}}$). These leakages in the form of varying diameter orifices would lead to a change in the mass flow. Other than $f_{W_{af}}$ (which is of medium severity), the other mass flow faults are of high severity, as they occur after the compressor and closer to the engine block (where the pressure is higher). Unattended mass flow faults in these engine components could cause a degradation in the engine performance (such as overpressure and increased emissions) as well as damages to the engine components themselves, especially if external abrasive particulate matters manage to enter the engine through the leakage orifices. As a result, this could lead to faults in other parts of the engine system, thus making efforts to isolate the original leakage fault more difficult.

The faults and their characteristics are summarized in Table \ref{tab:fault}.

\subsection{Fault Isolation Analysis from Model}
Using the differential equations in ``\nameref{AppB},'' a structural model of the engine system with the faults defined in Table \ref{tab:fault} is constructed. The structural model shows the relationships among the unknown variables of the engine system, the known variables (actuators and sensors), and the faults. The structural analysis is a useful tool for early determination of fault isolability, which shows how different levels of fault knowledge are incorporated into a structural fault isolability analysis and their results in different fault isolability conclusions \cite{ducstegor2006structural}.

\begin{figure}[t!]
\centering
\includegraphics[width=0.9\columnwidth]{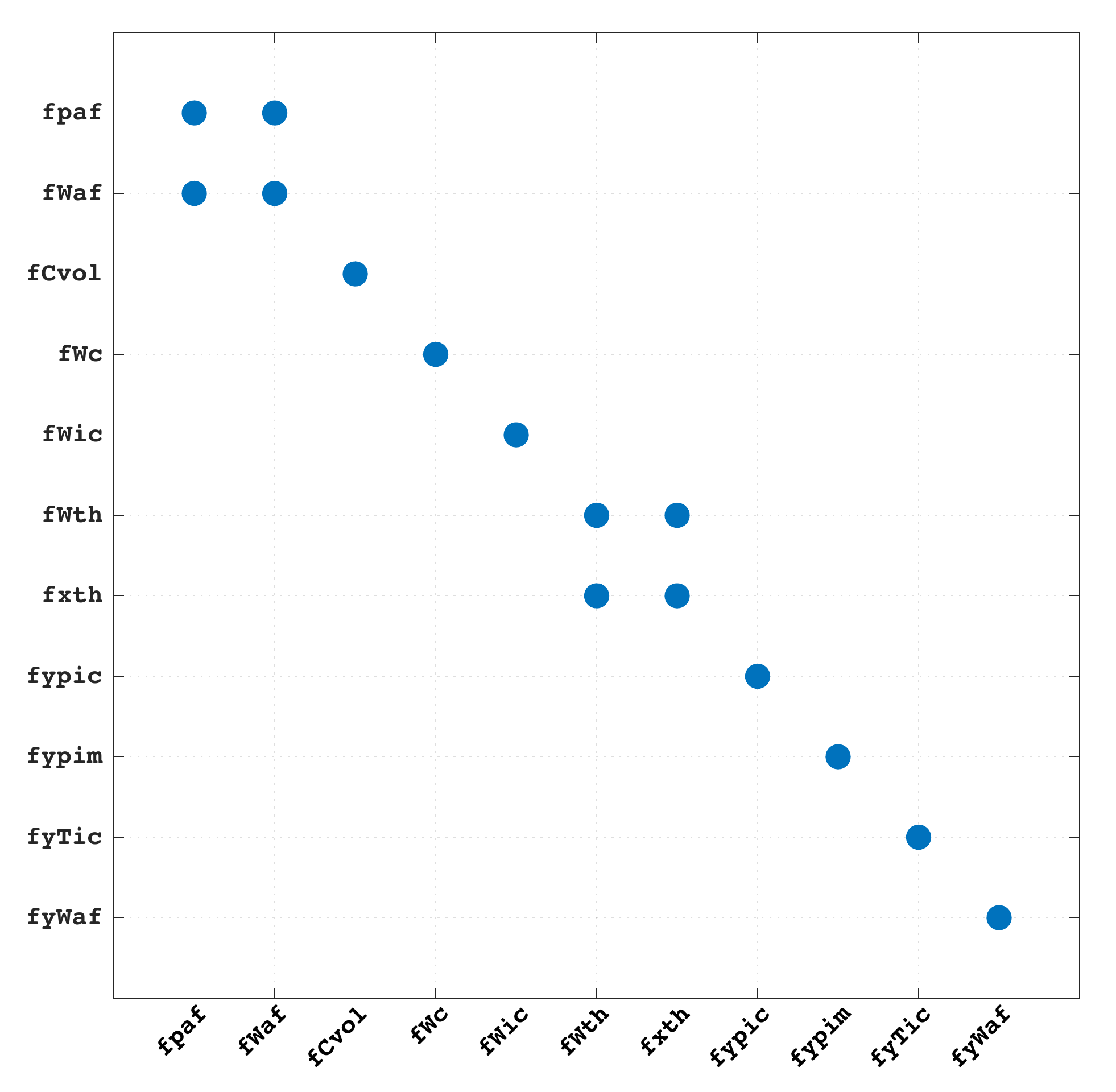}
\caption{Fault isolation matrix (FIM) constructed based on the structural model of the engine system. The figure shows that there are two pairs of nonisolable faults: $\{f_{p_{af}}, f_{W_{af}}\}$ and $\{f_{W_{th}}, f_{x_{th}}\}$. This is an ideal result for fault isolation, assuming no limits to the magnitudes and shapes of the faults.}
\label{fig:FIMstruc}
\end{figure}

Using the structural model, the fault isolation matrix (FIM) is then generated for initial fault isolation analysis. The FIM is a square matrix, where each row and column corresponds to a fault. A dot is placed at position ($i,j$) to indicate that fault $f_{i}$ is not isolable from fault $f_{j}$. Figure \ref{fig:FIMstruc} shows the FIM for the engine system given the current sensors setup. Two pairs of nonisolable faults can be observed; $f_{p_{af}}$ is not isolable from $f_{W_{af}}$, and $f_{W_{th}}$ is not isolable from $f_{x_{th}}$. However, this is a best-case performance of fault isolation in theory, as this method does not consider the magnitudes and shapes of the faults, model uncertainties, and disturbances. Thus, this method is not able to provide an accurate representation of the actual fault isolation capability. As such, this simulation testbed (with bounded magnitudes of the faults and consideration of sensor noise) provides a more realistic outlook on the fault isolability for the engine system. See ``\nameref{SMFIM}'' for a general tutorial on the structural model and FIM. Further information on the studies and development of the structural model and the FIM can be found in \cite{ducstegor2006structural}.

\section{Design and Generation of Residuals}\label{Residuals}
\subsection{Introduction}\label{ori7}
\begin{figure}[t!]
\begin{center}
\fbox{\input{./Figures/Figure7}}
    \caption{\label{BD}Block diagram representation of the closed-loop engine control system and the residuals generator. The subsystem within the blue dotted box is the closed-loop engine control system shown in Figure \ref{Engine}, while the subsystem within the red dashed box is the residuals generator. This figure also shows the locations where the faults are induced.}
\end{center}
\end{figure}
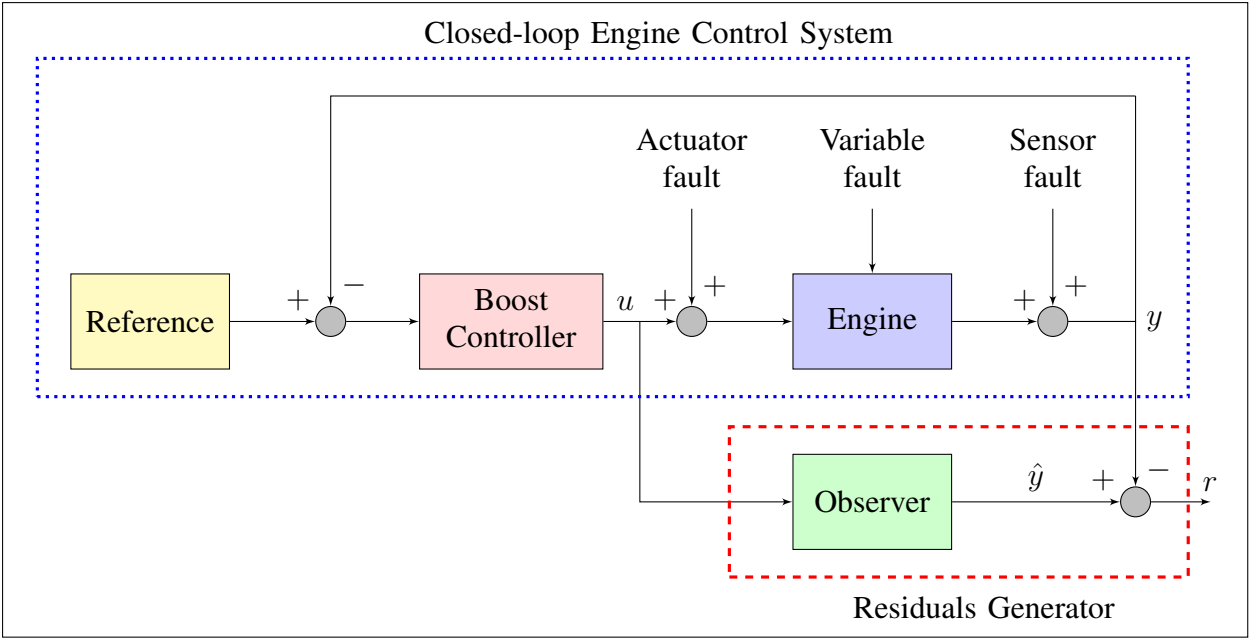

Initially, nine residuals are generated based on the sensors setup described in Table \ref{ActSen}. The simulation testbed is by default distributed with a state observer/estimator, which is constructed using the differential equations that describe the engine system. As such, the observer/estimator provides an estimate of the internal states of the engine system. The design of the observer can be replaced by other types of observers in the literature such as the sliding-mode observer (SMO) \cite{edwards2000sliding,Ng2012}, Kalman Filter \cite{simani2000diagnosis,kobayashi2003application}, or reduced-order observer \cite{darouach1996reduced,yang1998observer}. Therefore, this simulation testbed also enables researchers to design, develop, and compare strategies for residuals generator design for applications of state estimation and fault diagnosis of automotive engine systems. The residuals are generated by computing the difference between the sensor outputs of the model in Figure \ref{Engine} and the estimated outputs of an estimator/observer of the engine system: $r_{i} = \hat y_{i} - y_{i}$, where $\hat y_{i}$ and $y_{i}$ represent the  $i^{\mbox{\tiny th}}$ estimated and actual sensor outputs of the model, respectively. Figure \ref{BD} shows the overall block diagram representation of the closed-loop engine control system in Figure \ref{Engine} with the residuals generator. The residuals are then normalized using the standard deviation of the fault-free data as the measure of scale,
\begin{equation}
r_{i,N} = \frac{r_{i} - \mu_{r_{i}}}{\sigma_{r_{iNOMINAL}}},
\end{equation}
where $r_{i,N}$ is the normalized residual, $\mu_{r_{i}}$ the mean of the residual, and $\sigma_{r_{iNOMINAL}}$ the standard deviation of the corresponding residual during a fault-free scenario. These normalized residuals are called the “Original 9” and are listed in Table \ref{tab:residuals}.

\begin{table}[t!]
\caption{\label{tab:residuals}Default Residuals (“Original 9”) for fault detection given the sensors setup in Table \ref{ActSen}.}
\centering
\begin{tabularx}{0.7\textwidth}{lX} \hline
Residual		&	Description \\ \hline
$r_{T_{c}}$		& 	Residual for Compressor Temperature Sensor \\
$r_{p_{c}}$		& 	Residual for Compressor Pressure Sensor \\
$r_{T_{ic}}$	& 	Residual for Intercooler Temperature Sensor \\
$r_{p_{ic}}$	& 	Residual for Intercooler Pressure Sensor \\
$r_{T_{im}}$	& 	Residual for Intake Manifold Temperature Sensor \\
$r_{p_{im}}$	& 	Residual for Intake Manifold Pressure Sensor \\
$r_{W_{af}}$	& 	Residual for Air Filter Mass Flow Sensor \\
$r_{Tq_{e}}$	& 	Residual for Engine Torque Sensor \\
$r_{p_{em}}$	& 	Residual for Exhaust Manifold Pressure Sensor \\ \hline
\end{tabularx}
\end{table}

\begin{figure}[t!]
\centering
\includegraphics[width=0.98\columnwidth]{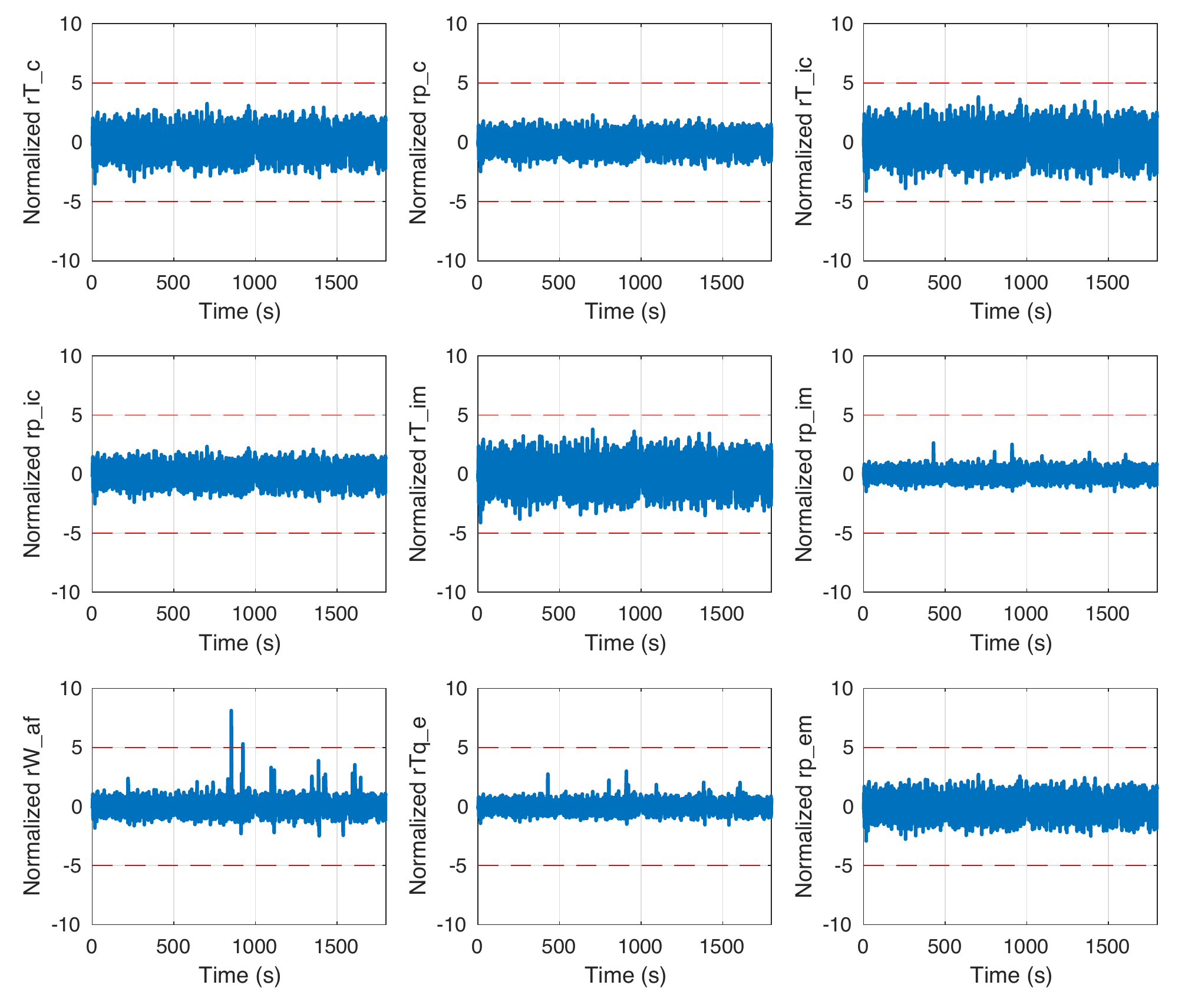}
\caption{Normalized plots of the “Original 9” for a fault-free scenario. All residuals have zero mean values, which indicate that both the model of the engine and the estimator produce almost identical outputs while being excited by the same control inputs when there are no faults in the engine system. The dashed lines are the fault detection thresholds to determine if the residuals have triggered and a fault has been detected.}
\label{fig:nofault}
\end{figure}

In a nominal fault-free scenario, all residuals have zero mean values. This indicates that both the model and the estimator of the engine produce similar actual and estimated outputs, respectively, while being excited by the same control inputs. Figure \ref{fig:nofault} shows the results of the residuals generated for a simulated fault-free scenario during the WLTP driving cycle profile. The dashed lines in Figure \ref{fig:nofault} represent the default fault detection threshold $J$, which determines if the residuals have triggered (that is, $|r_{i,N}| > J$), and hence indicates that a fault has been detected. For this simulation testbed, the threshold is tuned based on the nominal fault-free data to achieve a tradeoff between false-detection and missed detection rates. As such, the value of the threshold is initially set to $J = 5$. Of course, the value of the threshold can be easily changed.

\begin{figure}[t!]
\centering
\includegraphics[width=0.98\columnwidth]{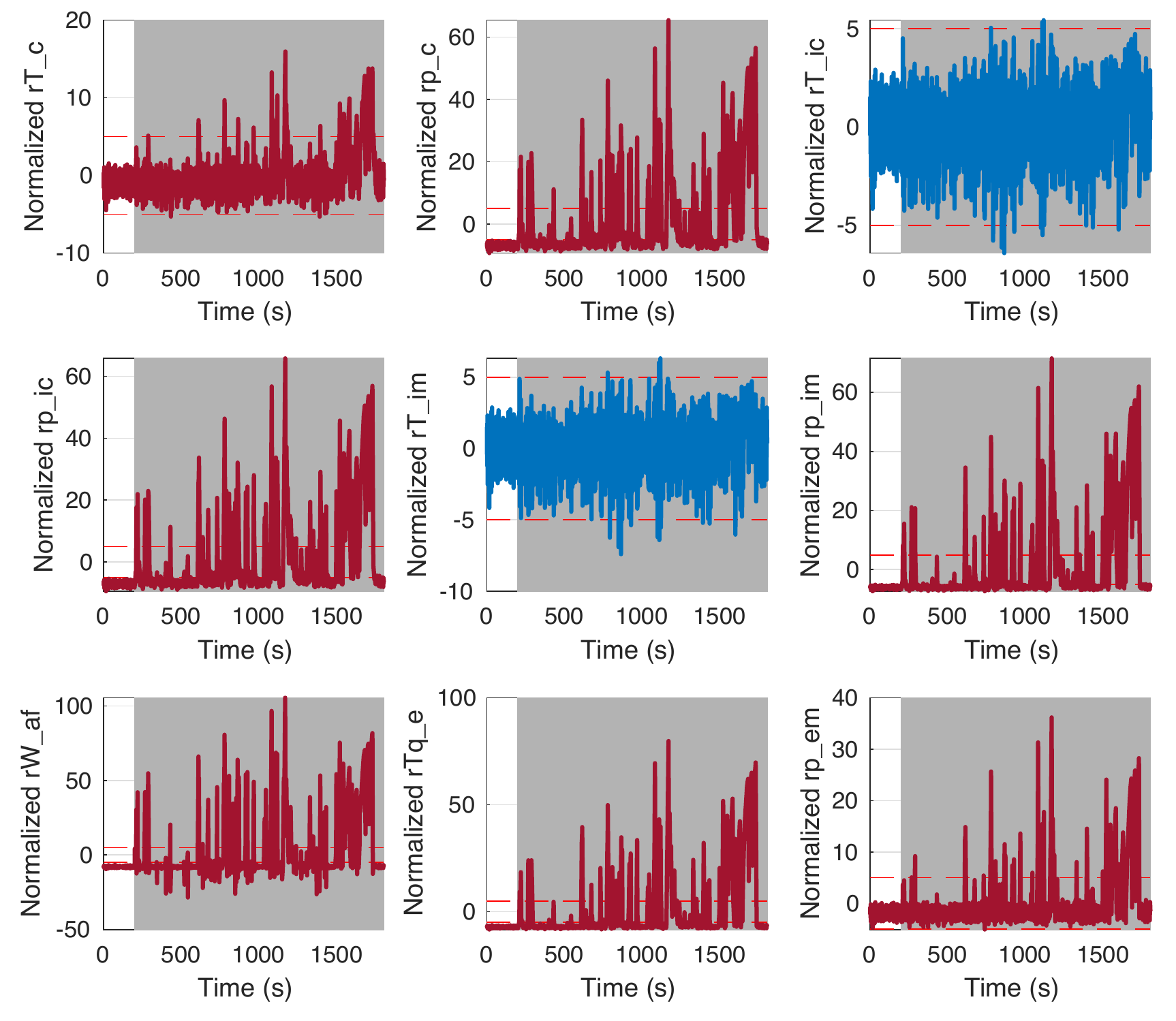}
\caption{Normalized plots of the “Original 9” for a loss of pressure in the air filter $f_{p_{af}}$. Plots in red are the residuals sensitive to the fault and hence triggered, while plots in blue are residuals that are not sensitive to the fault. This fault triggers all residuals except for $r_{T_{ic}}$ and $r_{T_{im}}$. The shaded regions show the duration for which the fault is active.}
\label{fig:simfpaf}
\end{figure}

\begin{figure}[t!]
\centering
\includegraphics[width=0.98\columnwidth]{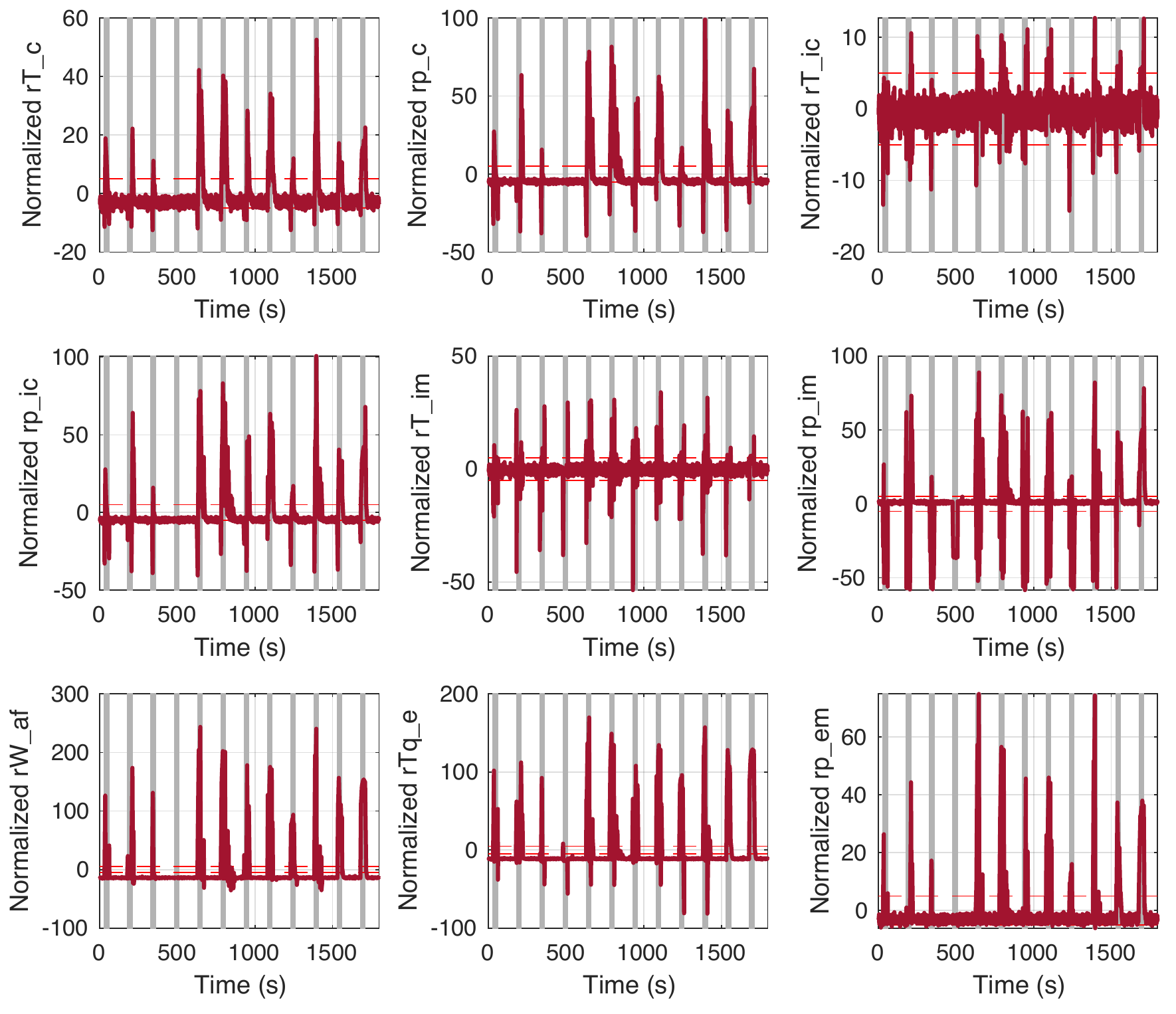}
\caption{Normalized plots of the “Original 9” for a stuck intake valve timing $f_{C_{vol}}$. This fault triggered all residuals. The shaded regions show the duration for which the fault is active.}
\label{fig:simfCvol}
\end{figure}

\begin{figure}[t!]
\centering
\includegraphics[width=0.98\columnwidth]{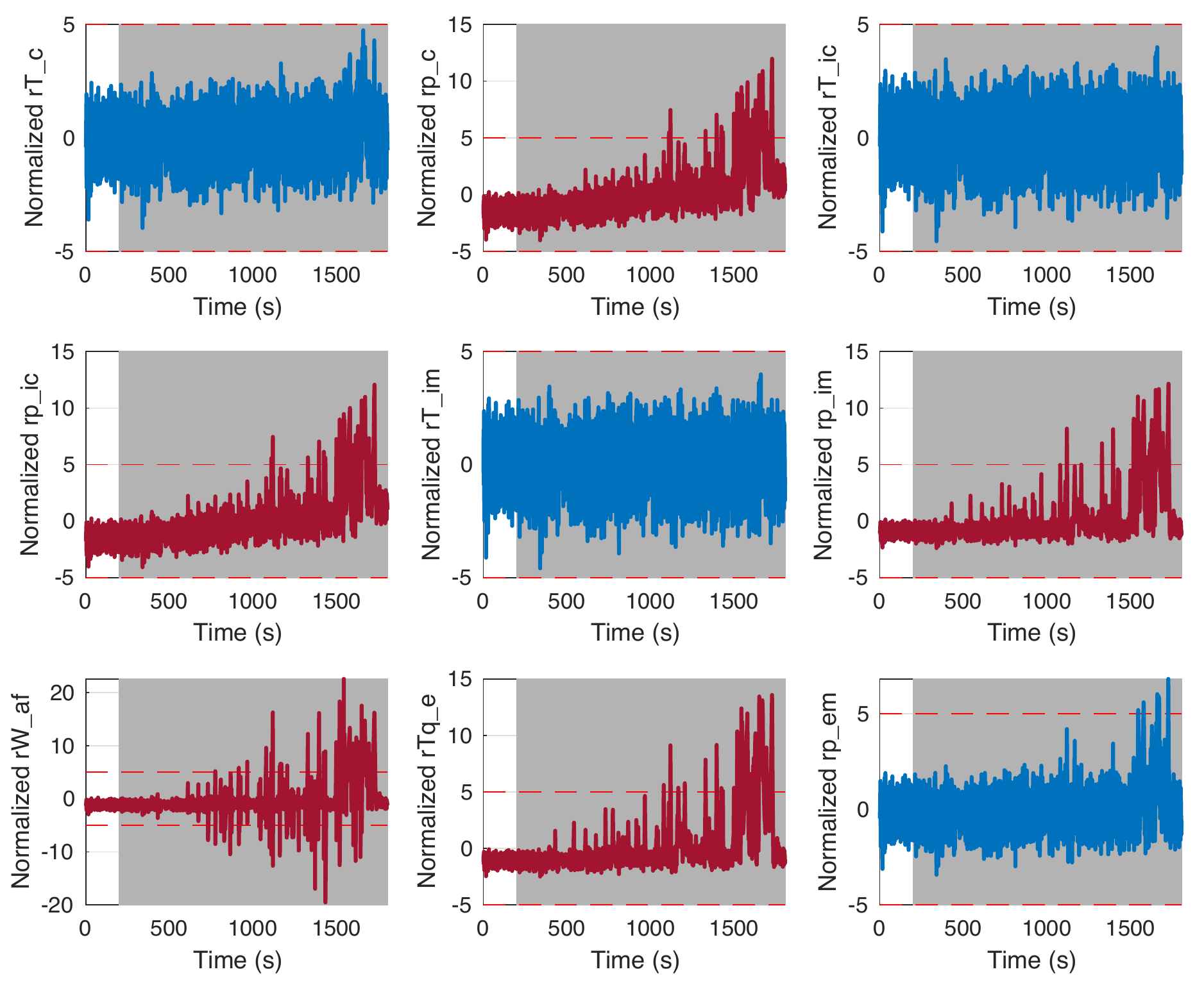}
\caption{Normalized plots of the “Original 9” for an air leakage between the air filter and the compressor $f_{W_{af}}$. Plots in red are the residuals sensitive to the fault and hence triggered, while plots in blue are residuals that are not sensitive to the fault. This fault triggered all residuals except for those related to temperature and $r_{p_{em}}$. The shaded regions show the duration for which the fault is active.}
\label{fig:simfWaf}
\end{figure}

\begin{figure}[t!]
\centering
\includegraphics[width=0.98\columnwidth]{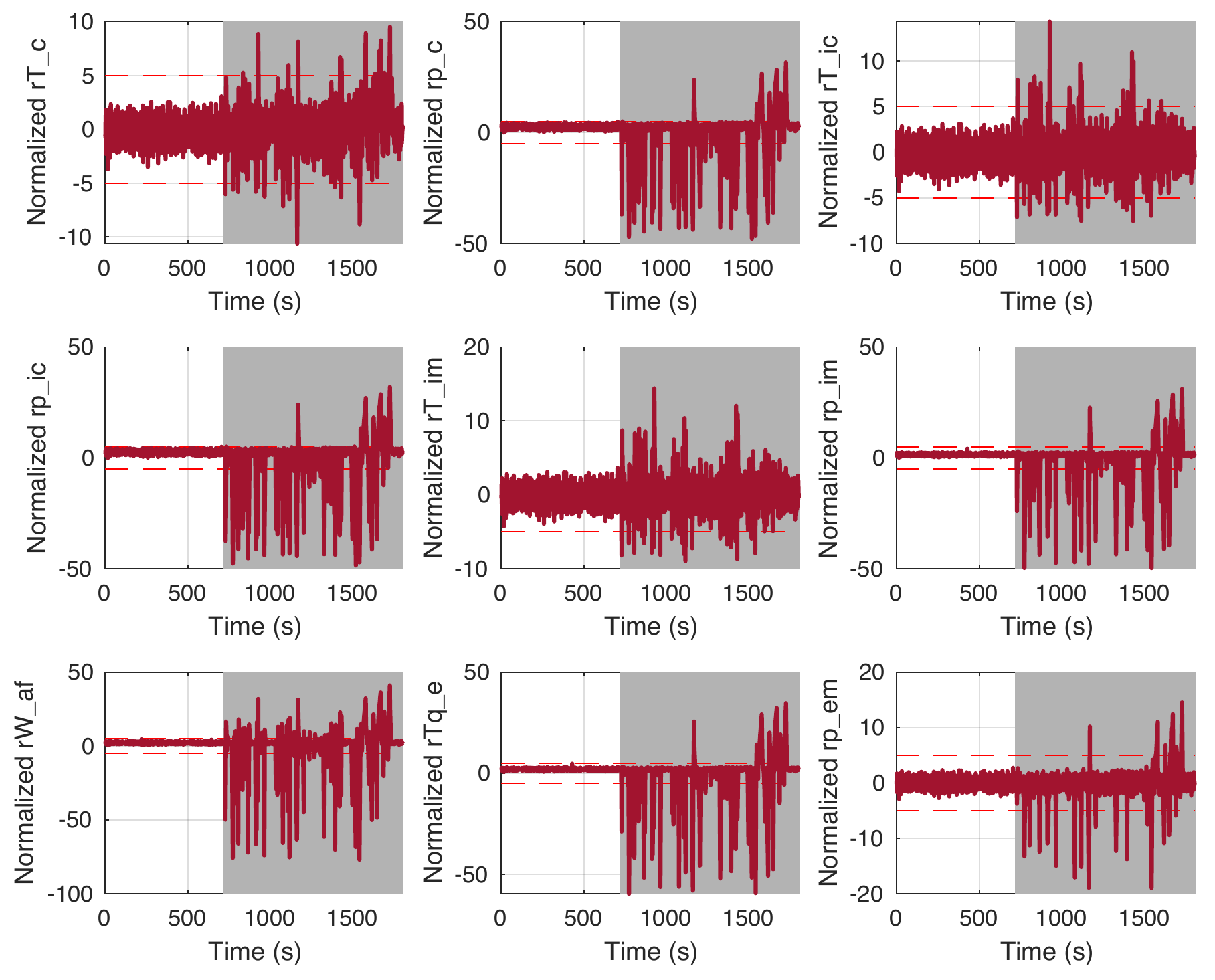}
\caption{Normalized plots of the “Original 9” for an air leakage between the compressor and the intercooler $f_{W_{c}}$. This fault triggered all residuals. The shaded regions show the duration for which the fault is active.}
\label{fig:simfWc}
\end{figure}

\begin{figure}[t!]
\centering
\includegraphics[width=0.98\columnwidth]{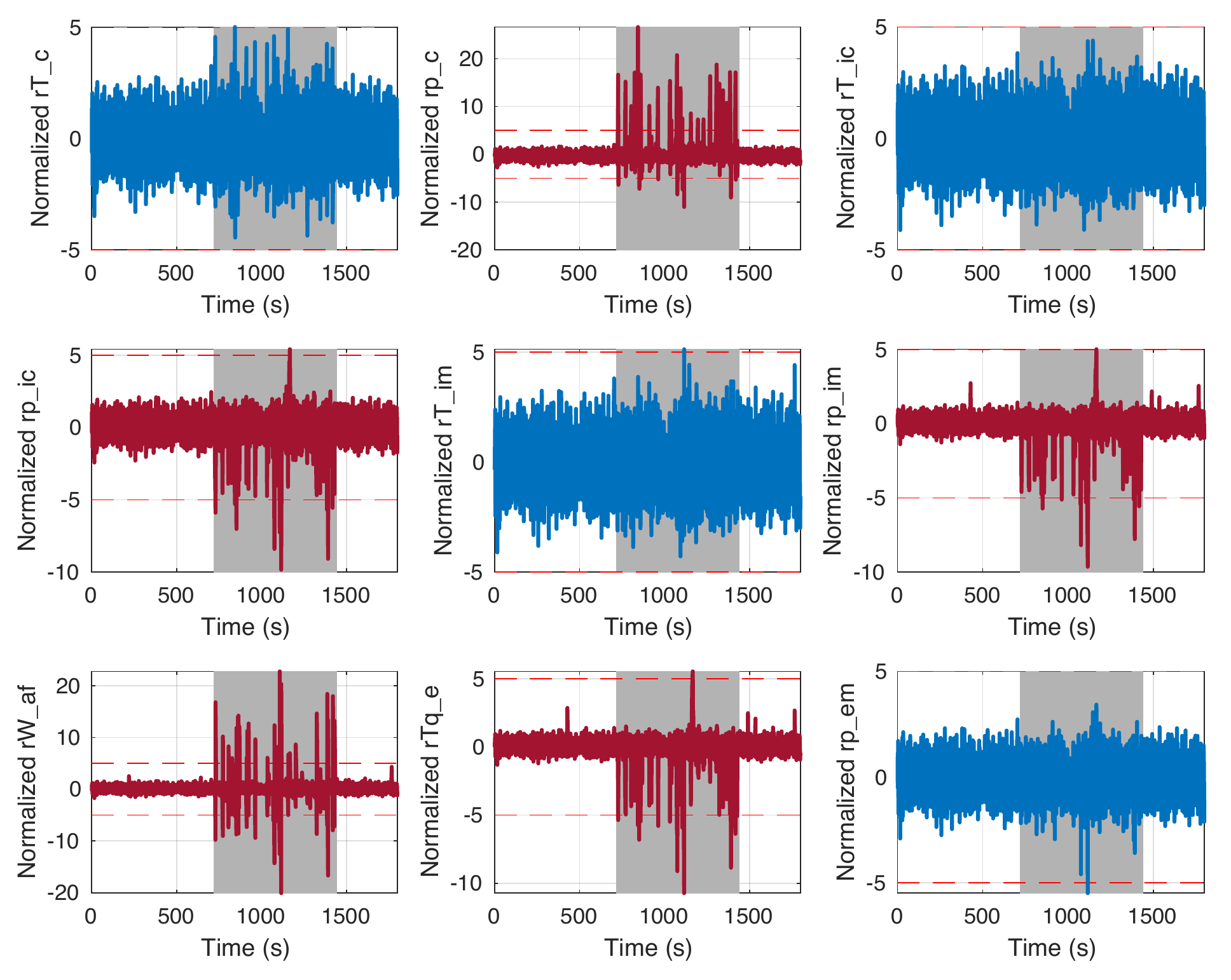}
\caption{Normalized plots of the “Original 9” for an air leakage between the intercooler and the throttle $f_{W_{ic}}$. Plots in red are the residuals sensitive to the fault and hence triggered, while plots in blue are residuals that are not sensitive to the fault. This fault triggered all residuals except for those related to temperature and $r_{p_{em}}$. The shaded regions show the duration for which the fault is active.}
\label{fig:simfWic}
\end{figure}

\begin{figure}[t!]
\centering
\includegraphics[width=0.98\columnwidth]{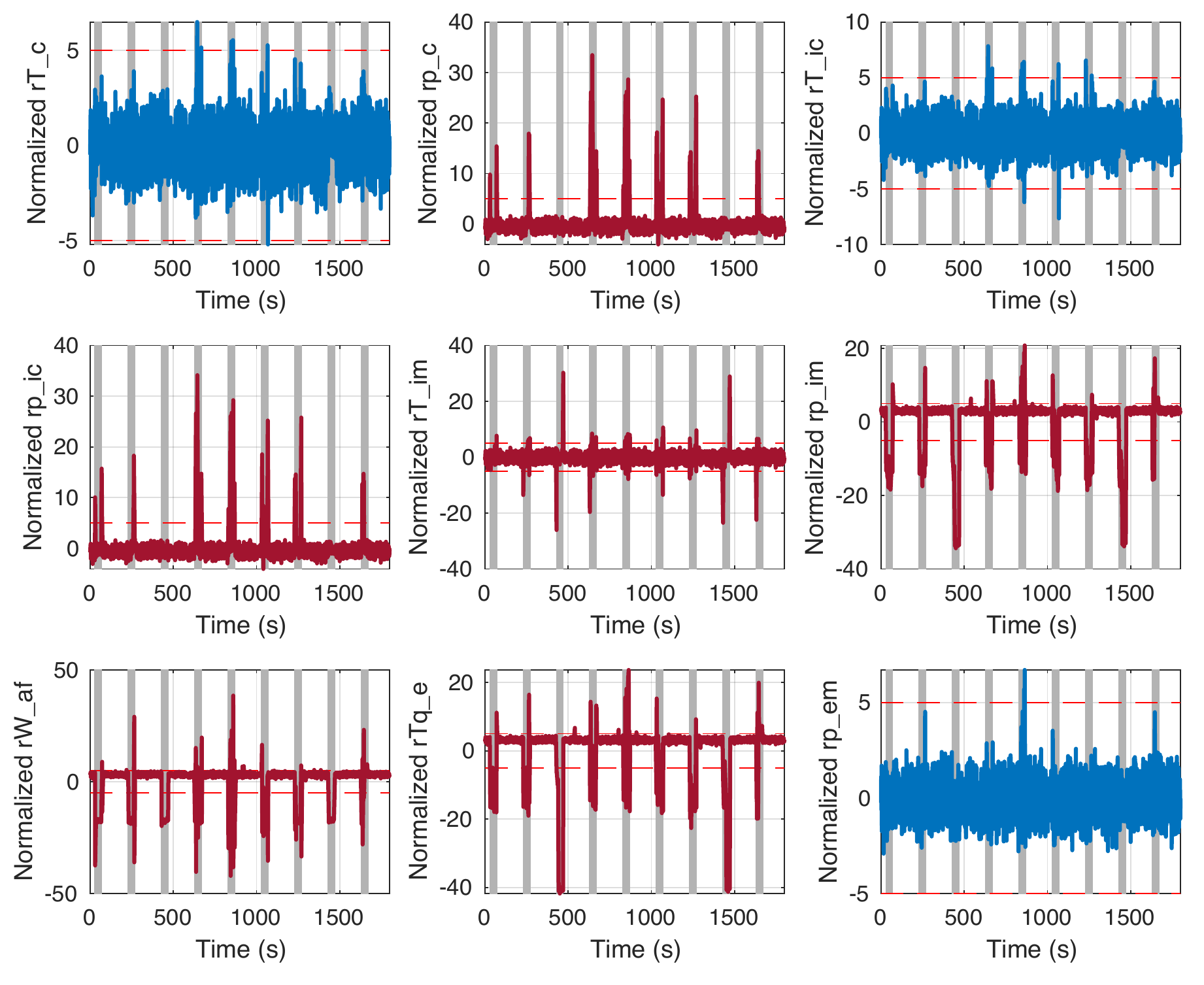}
\caption{Normalized plots of the “Original 9” for an air leakage after the throttle in the intake manifold $f_{W_{th}}$. Plots in red are the residuals sensitive to the fault and hence triggered, while plots in blue are residuals that are not sensitive to the fault. This fault triggered all residuals except for $r_{T_{c}}, r_{T_{ic}}$, and $r_{p_{em}}$. The shaded regions show the duration for which the fault is active.}
\label{fig:simfWth}
\end{figure}

\begin{figure}[t!]
\centering
\includegraphics[width=0.98\columnwidth]{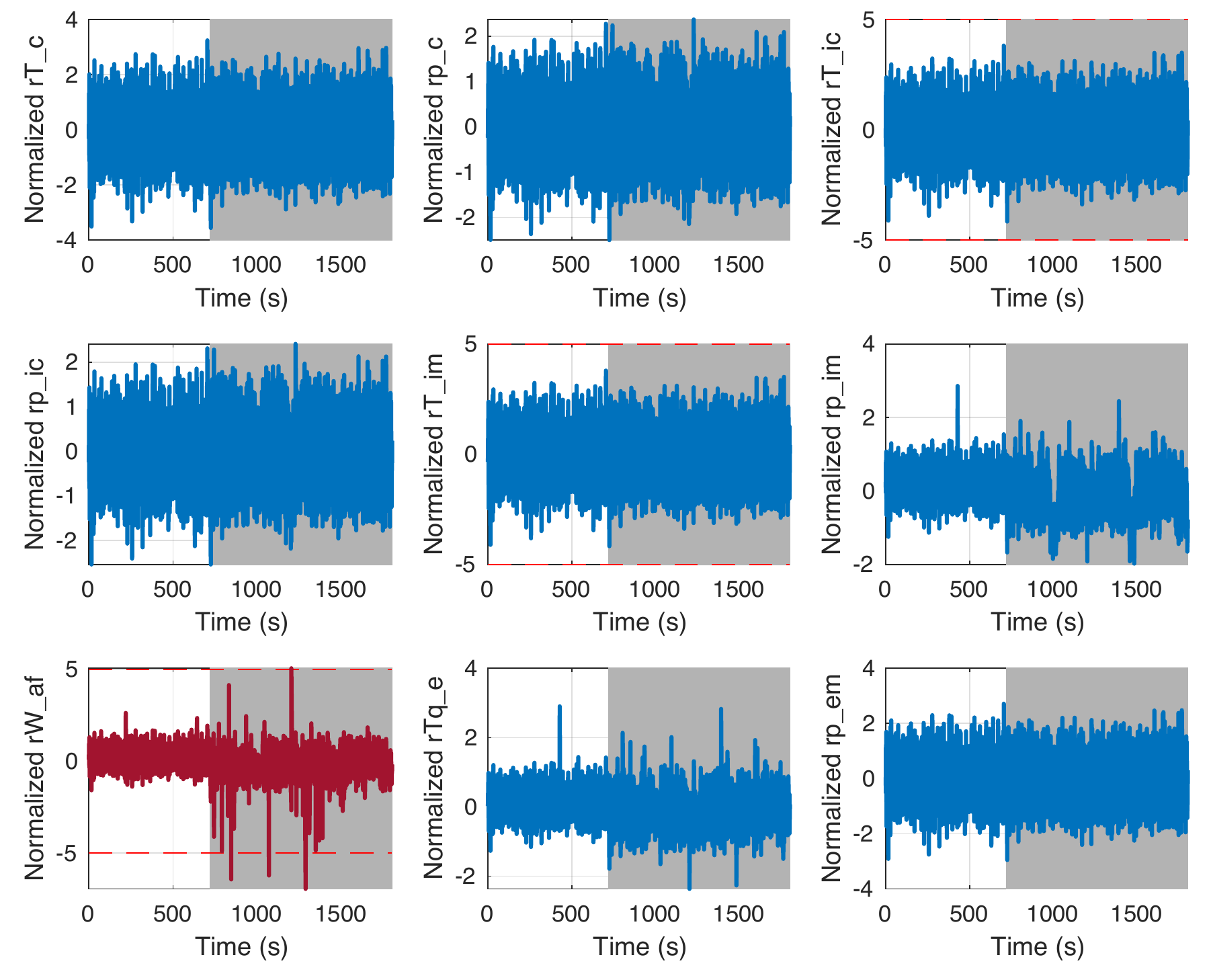}
\caption{Normalized plots of the “Original 9” for throttle position actuator error $f_{x_{th}}$. Plots in red are the residuals sensitive to the fault and hence triggered, while plots in blue are residuals that are not sensitive to the fault. This fault only triggered the $r_{W_{af}}$ residual. The shaded regions show the duration for which the fault is active.}
\label{fig:simfxth}
\end{figure}

\begin{figure}[t!]
\centering
\includegraphics[width=0.98\columnwidth]{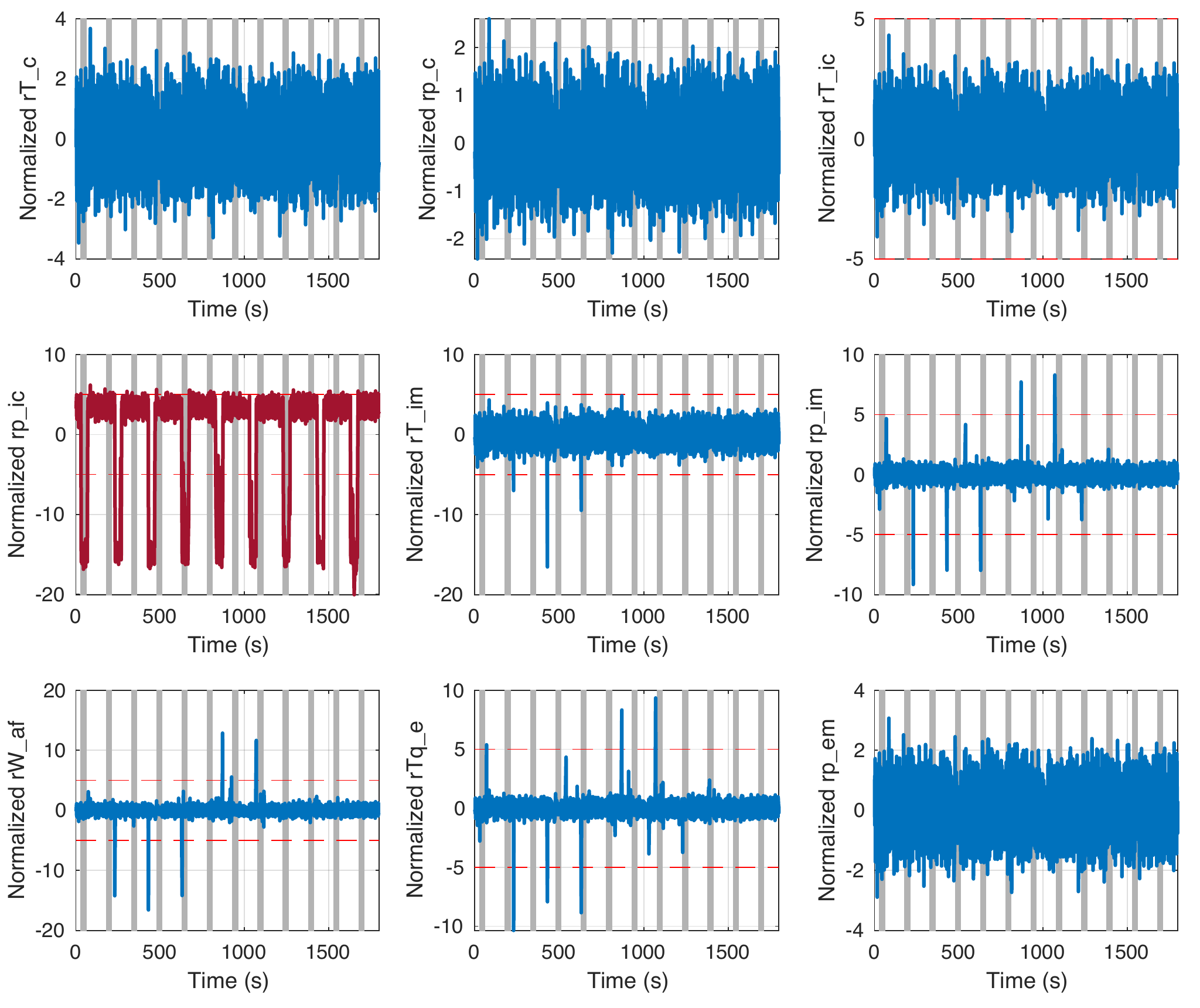}
\caption{Normalized plots of the “Original 9” for an intercooler pressure sensor fault $f_{y_{p_{ic}}}$. Plots in red are the residuals sensitive to the fault and hence triggered, while plots in blue are residuals that are not sensitive to the fault. This fault only triggered the $r_{p_{ic}}$ residual. The shaded regions show the duration for which the fault is active.}
\label{fig:simfypic}
\end{figure}

\begin{figure}[t!]
\centering
\includegraphics[width=0.98\columnwidth]{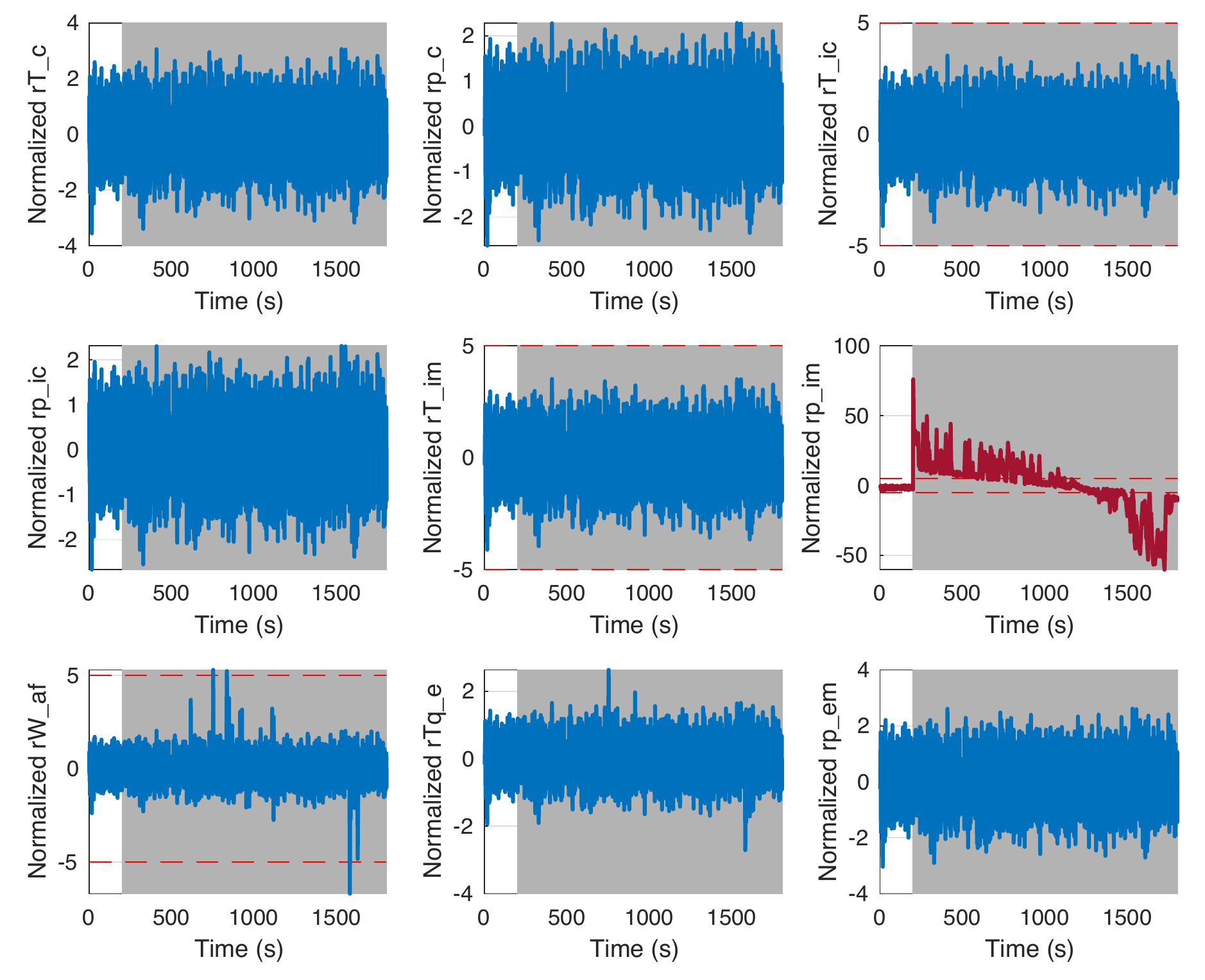}
\caption{Normalized plots of the “Original 9” for an intake manifold pressure sensor fault $f_{y_{p_{im}}}$. Plots in red are the residuals sensitive to the fault and hence triggered, while plots in blue are residuals that are not sensitive to the fault. This fault only triggered the $r_{p_{im}}$ residual. The shaded regions show the duration for which the fault is active.}
\label{fig:simfypim}
\end{figure}

\begin{figure}[t!]
\centering
\includegraphics[width=0.98\columnwidth]{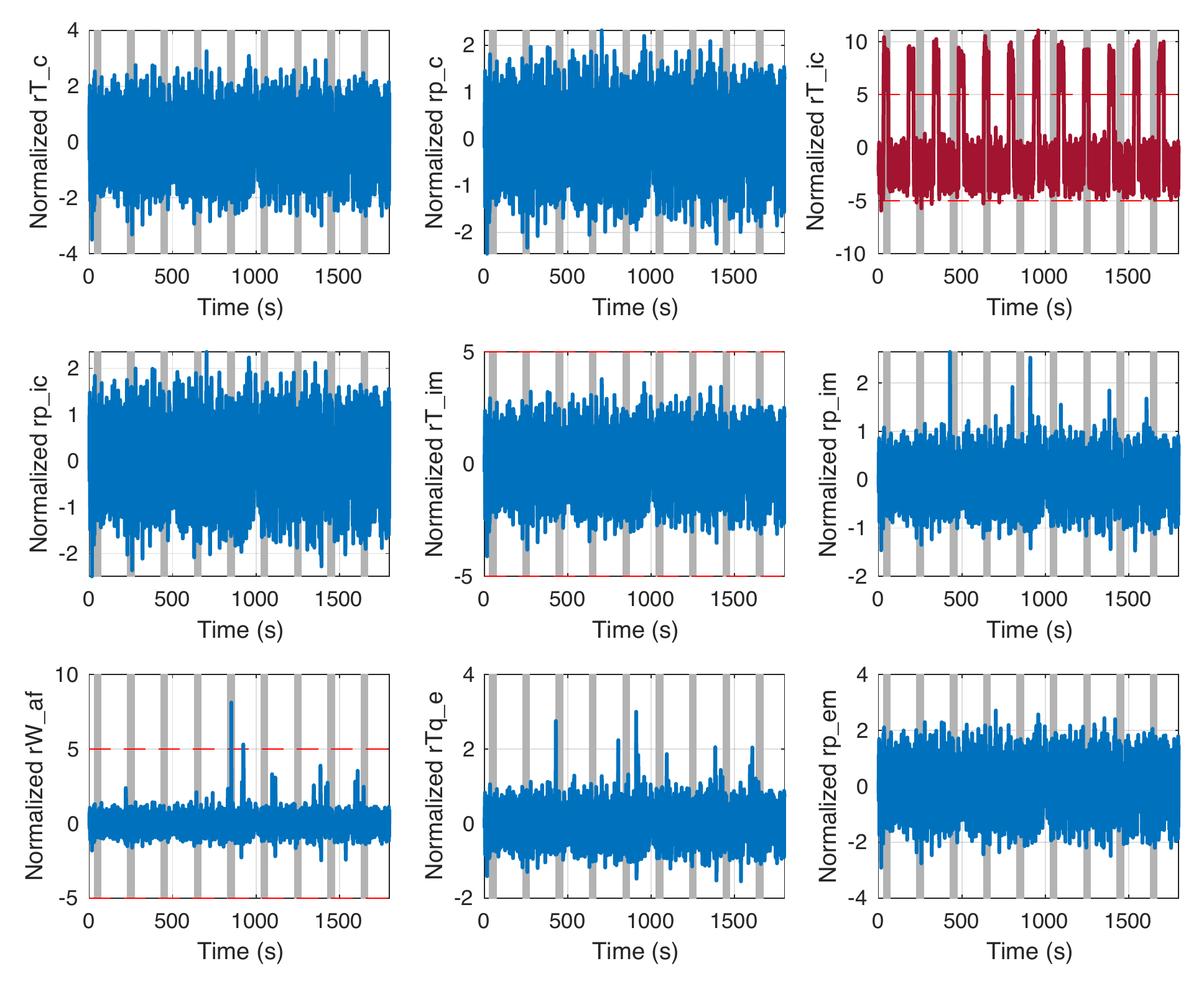}
\caption{Normalized plots of the “Original 9” for an intercooler temperature sensor fault $f_{y_{T_{ic}}}$. Plots in red are the residuals sensitive to the fault and hence triggered, while plots in blue are residuals that are not sensitive to the fault. This fault only triggered the $r_{T_{ic}}$ residual. The shaded regions show the duration for which the fault is active.}
\label{fig:simfyTic}
\end{figure}

\begin{figure}[t!]
\centering
\includegraphics[width=0.98\columnwidth]{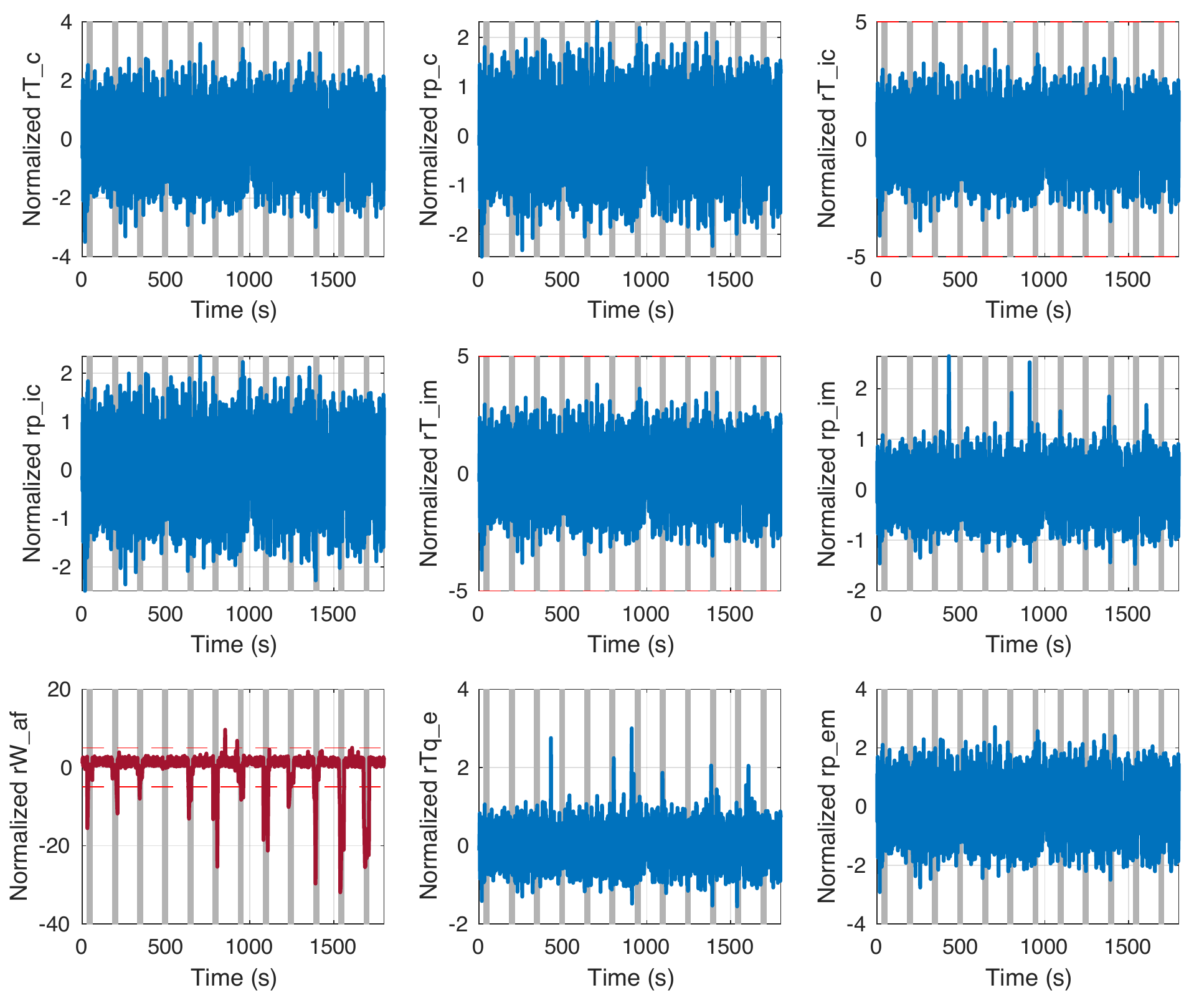}
\caption{Normalized plots of the “Original 9” for an air filter flow sensor fault $f_{y_{W_{af}}}$. Plots in red are the residuals sensitive to the fault and hence triggered, while plots in blue are residuals that are not sensitive to the fault. This fault only triggered the $r_{W_{af}}$ residual. The shaded regions show the duration for which the fault is active.}
\label{fig:simfyWaf}
\end{figure}

The engine control system and the residuals generator are then simulated with the faults in Table \ref{tab:fault}. Residuals that are sensitive to the corresponding faults would trigger and produce nonzero mean values. Figures \ref{fig:simfpaf}--\ref{fig:simfyWaf} show the simulation results for the “Original 9” residuals during the WLTP driving cycle profile when the engine system is induced with the faults. Only single-fault scenarios are currently considered. The figures show the dynamics of the engine system and the nature of the faults (that is, if they are actuator, sensor, or variable fault type, or if they are induced as an abrupt or incipient fault) influence the corresponding residuals that are sensitive to the individual faults to exceed the threshold and to trigger. For example, all sensor faults \{$f_{y_{p_{ic}}}, f_{y_{p_{im}}}, f_{y_{T_{ic}}}, f_{y_{W_{af}}}$\} triggered only one residual each, as they do not directly affect the states of the engine system (see Figures \ref{fig:simfypic}--\ref{fig:simfyWaf}). However, they could still affect the system indirectly if they are used as feedback signals. Since the actuator and variable faults directly affect the dynamics of the engine system, more residuals are sensitive to such faults. Therefore, if detected, it is usually easier to isolate sensor faults, compared to variable faults. By collectively identifying which residuals were triggered for the faults induced, fault isolation analysis to locate the fault in the engine system can then be performed.

\begin{figure}[t!]
\centering
\includegraphics[width=0.98\columnwidth]{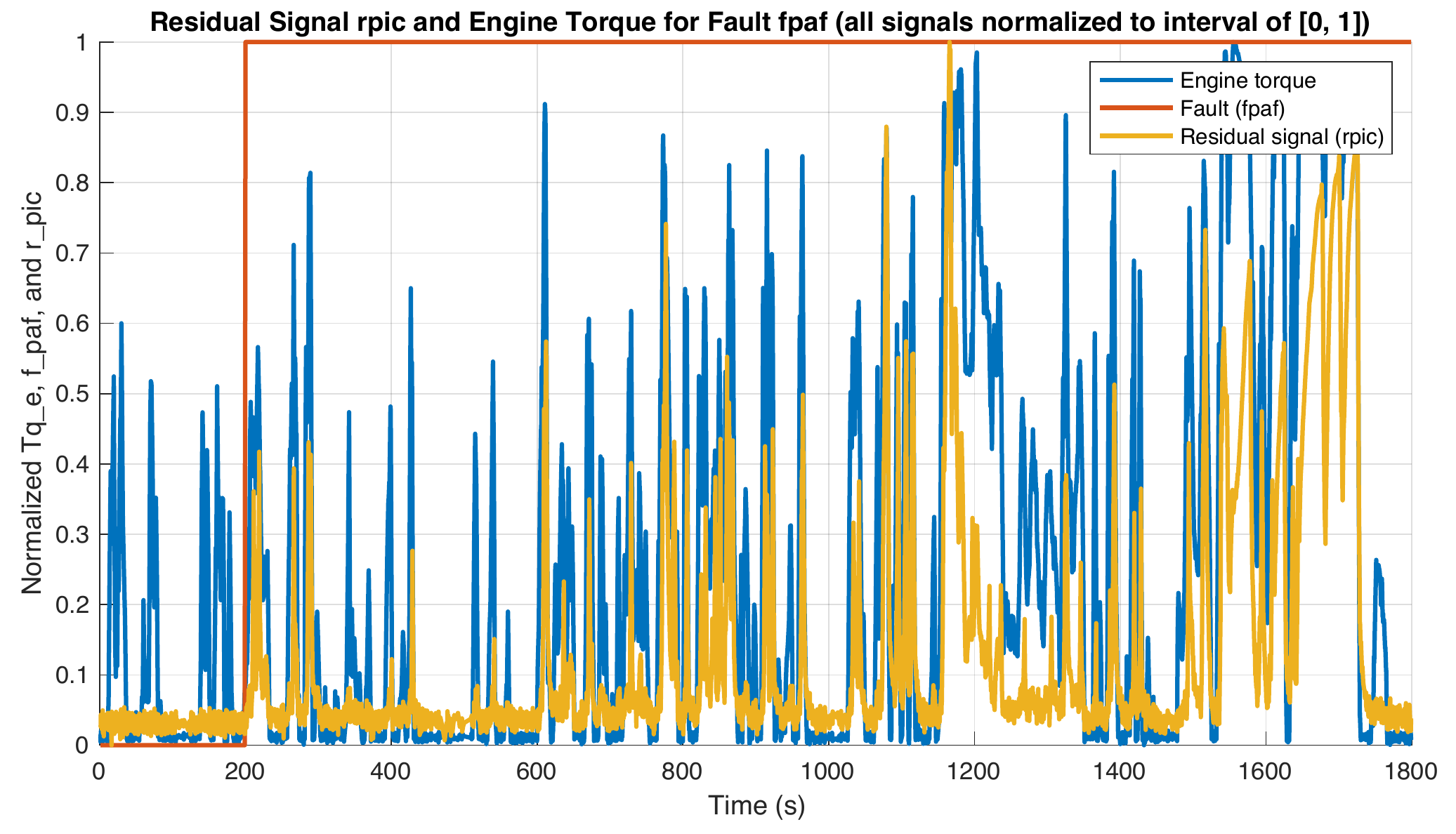}
\caption{The $r_{p_{ic}}$ residual signal generated (amber line) for a $f_{p_{af}}$ fault (red line) with the engine torque (blue line) during a Worldwide harmonized Light vehicles Test Procedures (WLTP) driving cycle profile.}
\label{fig:resTq}
\end{figure}

During simulations of faults in real-world conditions (especially for nonrepetitive driving cycles such as the WLTP), it is interesting to visualize the effects of the faults on the residuals. Figure \ref{fig:resTq} shows that for long-term or permanent faults (such as a gradual increase of restricted pressure in the air filter $f_{p_{af}}$), the residuals might exhibit occasional spikes. These spikes are influenced by the engine dynamics, such as an increase or decrease in the engine torque. This indicates that the amplitude of the faults and the engine dynamics would affect the outlook of the residuals generated and hence, they must be considered during the design of the fault diagnosis scheme.

\subsubsection{Fault Detection Requirements}
The suggested requirements for fault detection are as follows:
\begin{itemize}
\item \textit{Time for fault detection}: The decision for fault detection is made based on the amplitude of the residuals (that is, if the residuals exceed the threshold $J$) as well as the duration that they remain above the threshold. For this simulation testbed, the fault should be detected if the duration that the residuals exceed the threshold is $t_{f} > 3$ s.
\item \textit{Missed detections}: This testbed is designed such that the amplitudes of the faults are of sizes that they all should be detected.
\end{itemize}

\subsubsection{Fault Isolation Analysis from Simulations}
\begin{figure}[t!]
\centering
\includegraphics[width=0.9\columnwidth]{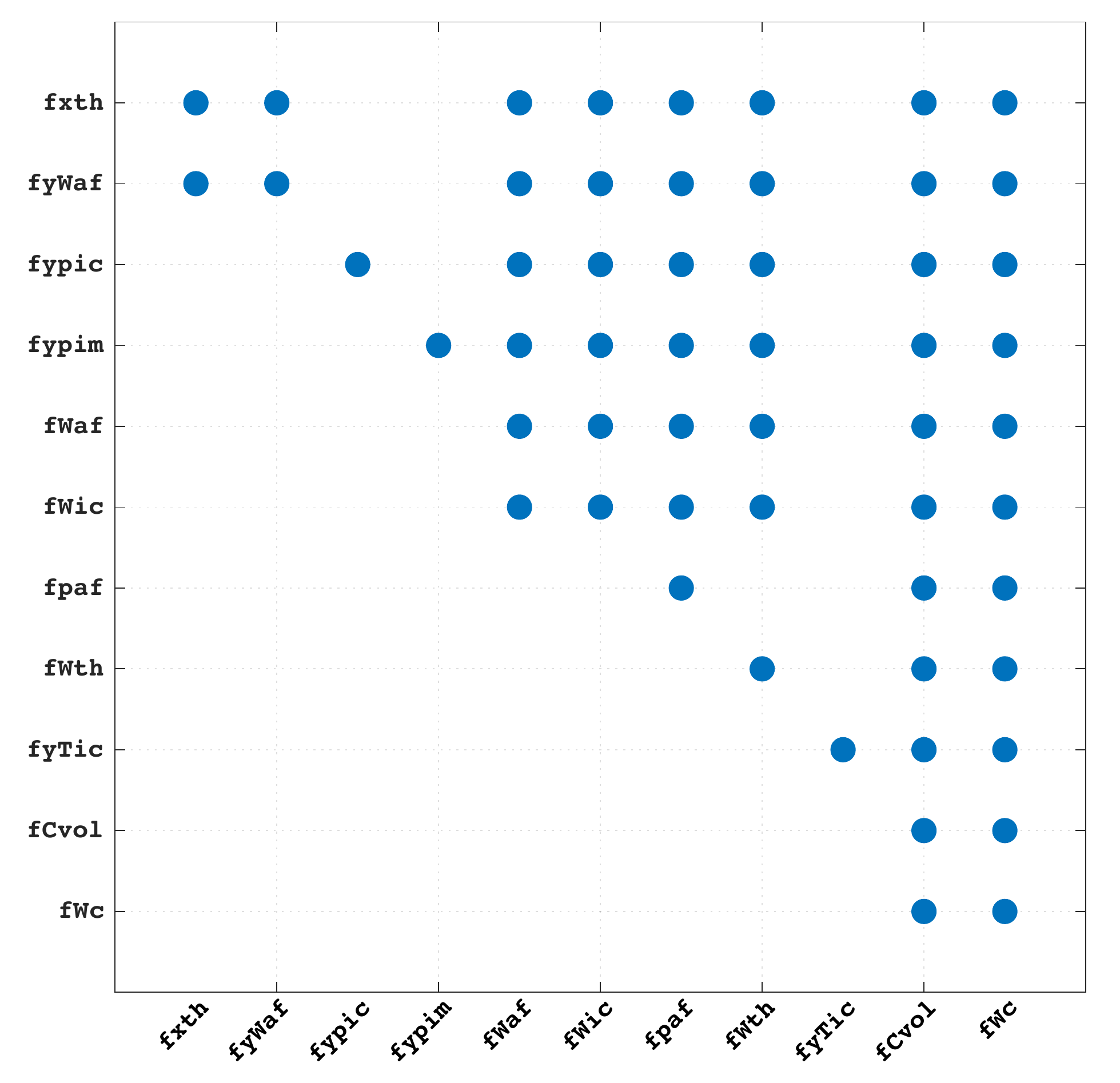}
\caption{Fault isolation matrix (FIM) constructed based on the fault sensitivity matrix (FSM) in Table \ref{FSMdef}. This is a more realistic representation on the fault isolation analysis, as it considers the magnitudes and shapes of the faults.}
\label{fig:FIMsimu}
\end{figure}

The fault sensitivity matrix (FSM) in Table \ref{FSMdef} can be constructed from the simulation results shown in Figures \ref{fig:simfpaf}--\ref{fig:simfyWaf}. The FSM is tabulated by placing a value of `1' if the residual is triggered by the specific fault and `0' otherwise. Using the FSM in Table \ref{FSMdef}, the FIM of the system for the current residuals design can then be constructed. Figure \ref{fig:FIMsimu} shows the FIM with a more realistic fault isolation performance of the system when the magnitudes and shapes of the faults acting on the engine system are also considered. See Figure \ref{fig:FIMstruc} for comparison. However, the results are not exciting, as many faults are not isolable from each other. Therefore, this model would serve as an excellent platform for designers and researchers to design and perform model-in-the-loop tests of fault diagnosis schemes, with application to actual automotive engine systems.

\begin{table*}[t!]
\caption{\label{FSMdef}The fault sensitivity matrix (FSM) of the “Original 9” residuals.}
\centering
\begin{tabular}{lccccccccccc} \hline
Residual & $f_{p_{af}}$	& $f_{C_{vol}}$	 & $f_{W_{af}}$ & $f_{W_{c}}$ & $f_{W_{ic}}$ &	$f_{W_{th}}$ & $f_{x_{th}}$ & $f_{y_{p_{ic}}}$ & $f_{y_{p_{im}}}$ & $f_{y_{T_{ic}}}$ &	$f_{y_{W_{af}}}$ \\ \hline
$r_{T_{c}}$ 	& 1	& 1 & 0	& 1	& 0	& 0	& 0	& 0	& 0	& 0 & 0 \\
$r_{p_{c}}$		& 1	& 1	& 1	& 1	& 1	& 1	& 0	& 0	& 0 & 0	& 0 \\
$r_{T_{ic}}$	& 0	& 1	& 0	& 1	& 0	& 0	& 0	& 0	& 0 & 1	& 0 \\
$r_{p_{ic}}$	& 1	& 1	& 1	& 1	& 1	& 1	& 0	& 1	& 0	& 0	& 0 \\
$r_{T_{im}}$	& 0	& 1	& 0	& 1	& 0	& 1	& 0	& 0	& 0 & 0 & 0 \\
$r_{p_{im}}$	& 1	& 1	& 1	& 1	& 1	& 1	& 0	& 0	& 1 & 0	& 0 \\
$r_{W_{af}}$	& 1	& 1	& 1	& 1 & 1	& 1	& 1	& 0	& 0	& 0 & 1 \\
$r_{Tq_{e}}$	& 1	& 1	& 1	& 1 & 1	& 1	& 0	& 0	& 0	& 0 & 0 \\
$r_{p_{em}}$	& 1	& 1	& 0	& 1 & 0	& 0	& 0	& 0	& 0	& 0 & 0 \\ \hline
\end{tabular}
\end{table*}

\section{The Simulation Environment}
\label{Simulation}
\begin{figure}[t!]
\begin{center}
\includegraphics[width=\textwidth]{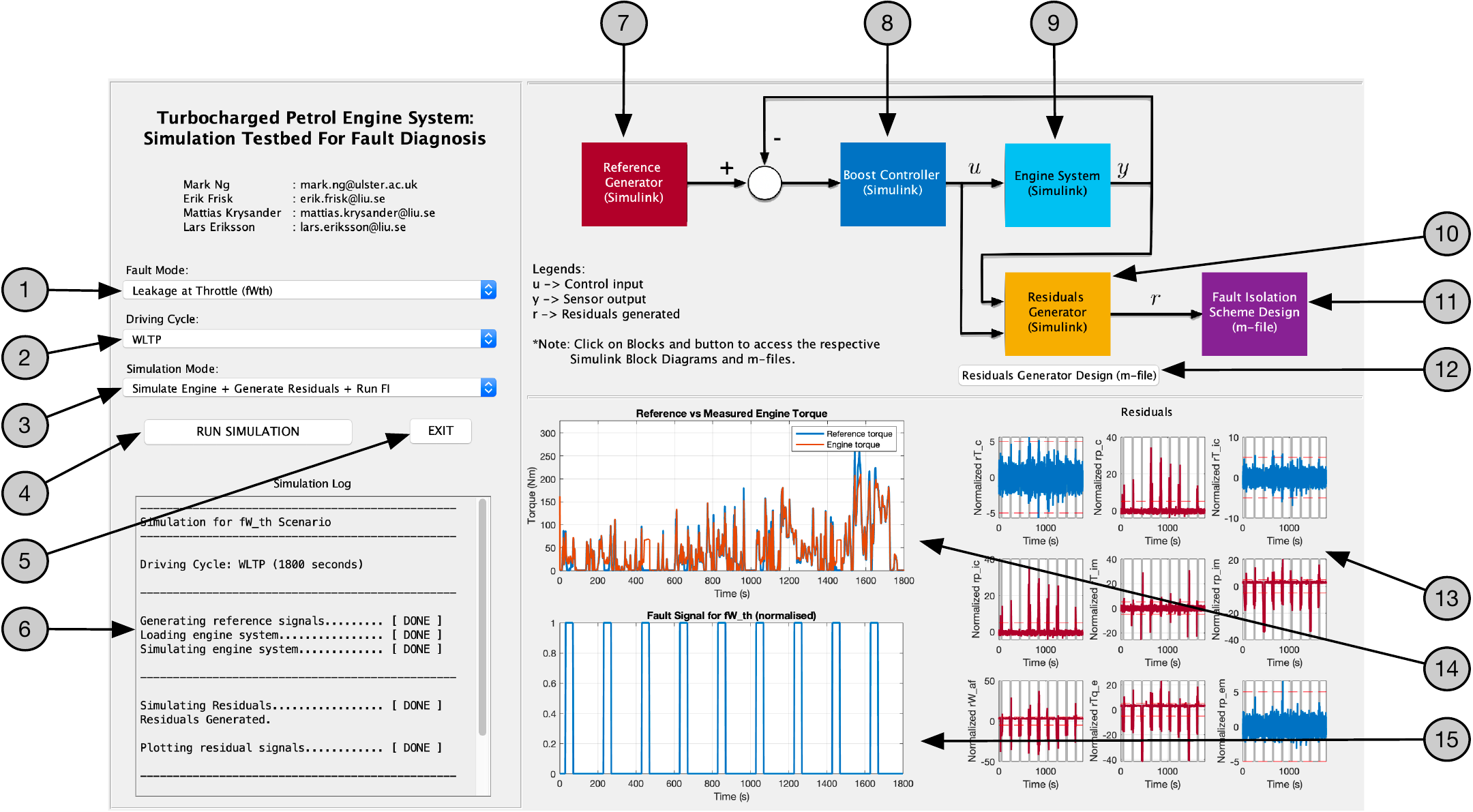}
\caption{\label{fig:GUI}The main GUI of the simulation testbed in Matlab; 1) Sets the fault mode for simulation. 2) Sets the driving cycle. 3) Sets the simulation mode. 4) Runs the simulation. 5) Exits and closes the testbed GUI. 6) Shows the simulation progress and log. 7) Select to open the reference generator Simulink model. 8--9) Select to open the boost controller and engine Simulink model. 10) Select to open the residuals generator Simulink model. 11) Select to open the fault diagnosis design scheme m-file. 12) Select to open the residuals generator design scheme m-file. 13) Displays the residuals generated. 14) Displays the reference torque and the actual torque of the engine. 15) Displays the fault signal induced (normalized).}
\end{center}
\end{figure}

Figure \ref{fig:GUI} shows the GUI of the simulation testbed in Matlab. Through this interface, the user can set the preferences for simulation settings, design, and test their residuals generation and fault diagnosis schemes, as well as view simulation results.

\subsection{Establishing Simulation Settings}
In the left section of the GUI are pop-up menus for the user to establish key simulation settings. The simulation settings available include
\begin{itemize}
\item {\it Fault Mode}: To induce any of the 11 faults defined in Table \ref{tab:fault}. A fault-free scenario is also available and is selected by default. As of current development, only single-fault scenarios are available.
\item {\it Driving Cycle}: A selection of four industrial-standard driving cycles (WLTP, NEDC, EUDC, and FTP-75).
\item {\it Simulation Mode}: A choice of two simulation modes, which either simulate only the engine for the chosen driving cycle or extend the simulation to also include generation of the residuals and the execution of the fault diagnosis algorithm. The latter choice would require design and coding inputs from the user.
\end{itemize}

\subsection{Design and Testing of Residuals Generation and Fault Diagnosis Schemes}
In the top-right section of the GUI, a block diagram representation of the engine control system, residuals generator, and fault diagnosis scheme can be found. The user can select each block to access the corresponding Simulink model or m-file. For example, the user could use the ”Residuals Generator (Simulink),” ”Residuals Generator Design (m-file),” and ”Fault Isolation Scheme Design (m-file)” components to edit their design and codes for the residuals generation and fault diagnosis algorithms. The ”RUN SIMULATION” pushbutton starts the simulation, while the ”EXIT” pushbutton exits the simulation environment and closes the GUI.

\subsection{Simulation Results}
The results obtained from the simulation are displayed in the bottom-right section of the GUI. The results displayed are the reference and the actual engine torques, as well as the normalized plot of the fault induced. A “Simulation Log” is also available in the bottom-left section of the GUI to show a summary of the simulation settings and provide an update in real time on the progress of the simulation. The plots and the “Simulation Log” are automatically saved into the folder {\tt /Results/\textit{DrivingCycle\_FaultMode\_Date}}, which is located in the same directory as the simulation files. A Matlab MAT-file containing key variables and data from the simulation is also saved (see Table \ref{tab:data}). Depending on the user's requirements, additional plots can be generated and saved into the same folder using the {\tt SavePlot()}command, and additional messages can be displayed onto the “Simulation Log” using the {\tt PrintLog()}command.
\begin{table}[t]
\caption{The variables saved into the MAT-file after each simulation run. The user could then use these data for further processing and analysis towards the design of the fault diagnosis scheme.}
\begin{center}
\begin{tabular}{lp{0.8\textwidth}} \hline
Saved Variable & Description \\ \hline
\texttt{omega\_eREF\_sync}				& Reference engine speed, $\omega_{eREF}$ \\
\texttt{Tq\_eREF\_sync} 				& Reference engine torque, $Tq_{eREF}$ \\
\texttt{inputSig\_sync}					& 5 actuator measurements of the engine ($A_{th}, u_{wg}, \omega_{eREF}, p_{amb}, T_{amb}$) \\
\texttt{outputSig\_sync} 				& 9 sensor measurements from the engine \newline($T_{c}, p_{c}, T_{ic}, p_{ic}, T_{im}, p_{im}, p_{em}, W_{af}, Tq_{e}$) \\
\texttt{statesSig\_sync} 				& 13 states of the engine ($T_{af}, p_{af}, T_{c}, p_{c}, T_{ic}, p_{ic}, T_{im}, p_{im}, T_{em}, p_{em}, T_{t}, p_{t}, \omega_{t}$) \\
\texttt{faultSig\_sync} 				& Normalized data of the faults (the selected induced fault would have nonzero data, except when “Fault-free” scenario is selected where all faults would have data of value zero) \\
\texttt{residualSig\_sync} 				& Data for all “Original 9” residuals based on the current sensors setup ($r_{T_c}, r_{p_c}, r_{T_{ic}}, r_{p_{ic}}, r_{T_{im}}, r_{p_{im}}, r_{W_{af}}, r_{Tq_e}, r_{p_{em}}$). Note that these data are only generated if Simulation Mode 2 is selected. \\ \hline
\end{tabular}
\end{center}
\label{tab:data}
\end{table}

\subsection{The Simulation Kit}\label{sec:Kit}
The simulation kit is available as opensource and can be downloaded from \url{https://github.com/nkymark/TCSISimTestbed}. The simulation kit contains the following key files:
\begin{itemize}
	\item {\tt main.m} — Main execution file. Run this file to start the GUI.
	\item {\tt Engine.mdl} — Simulink model of the closed-loop nonlinear engine system shown. Open the model from the GUI using either the “Boost Controller (Simulink)” or “Engine System (Simulink)” blocks.
	\item {\tt GenerateResiduals.m} — Codes for the residuals generation algorithm to be placed here. Open the file from the GUI using the “Residuals Generator Design (m-file)” button.
	\item {\tt ResidualsGen.mdl} — Simulink model of the residuals generator. The model is called and run from {\tt GenerateResiduals.m}. The default residuals generated are also filtered and normalized, and with added signal noise. Open the model from the GUI using the “Residuals Generator (Simulink)” block. Replace the “Residuals Generator” in the Simulink model as desired to accommodate other methods for residuals generation.
	\item {\tt RunFI.m} — Algorithm for fault diagnosis to be placed here. Open the file from the GUI using the “Fault Isolation Scheme Design (m-file)” block.
\end{itemize}

\section{Conclusion} \label{Conclusion}
This article presented a simulation testbed for testing and evaluation of residuals generation and fault diagnosis schemes in a TCSI petrol engine system. Key features of the simulation testbed were emphasized, including: 1) a realistic nonlinear model of the engine system compared with the actual physical test bench; 2) the testbed enables researchers to simulate actuator, sensor, and variable faults in various components of the engine system without having to physically modify the engine test bench; 3) researchers are able to compare the performance of their fault diagnosis schemes against the presented structural model and FIM benchmark; 4) general simulation and fault settings can be easily configured using the GUI interface, and the testbed can be modified and is customizable to accommodate the different residuals generation as well as fault diagnosis schemes; and 5) the simulation kit is available as opensource and can be downloaded for research and/or teaching purposes. The data generated from the simulation testbed are suitable for the study of both model-based and data-driven fault diagnosis methods. This testbed will serve as an excellent platform to demonstrate the effectiveness in designing, simulating, and analyzing fault diagnostic schemes on automotive systems for the development and comparison of current and future research methods as well as for teaching initiatives.

Future developments of the simulation testbed include the activation of intermittent residuals to mimic actual applications where some residuals are turned off during certain driving conditions (such as rough terrains or extreme weather), so that they do not trigger a false alarm. The addition of faults in other parts of the engine and new simulation options such as weather will also be considered.

Some of the interesting research challenges in this field of study include, but certainly are not limited to: 1) the issue of robustness will always be one of the main and critical problems of any control systems, and since most automakers sell their vehicles all over the world, it is very difficult for one fault diagnosis method to remain robust against a variety of terrains, weather, driving styles, and traffic conditions; 2) with ever-increasing development of autonomous vehicles, it is important for systems to be aware of their health and perform self-diagnosis or self-healing to ensure occupants' lives are protected at all times; and 3) a combination of model-based and data-driven methods could enhance fault diagnosis performance, especially when combined with cloud-based technologies where a fleet of vehicles contribute data to a general pool in the cloud.

\section{Acknowledgment}
This research was supported by Volvo Car Corporation in Gothenburg, Sweden.

\newpage
\bibliographystyle{IEEETran}
\bibliography{ref}

\processdelayedfloats 

\sidebars 

\clearpage
\section[Summary]{Sidebar: Summary}
\label{sec:Summary}
Research on fault diagnosis on highly nonlinear dynamic systems such as the engine of a vehicle have garnered huge interest in recent years, especially with the automotive industry heading towards self-driving technologies. This article presents a novel opensource simulation testbed of a turbocharged spark ignited (TCSI) petrol engine system for testing and evaluation of residuals generation and fault diagnosis methods. Designed and developed using Matlab/Simulink, the user interacts with the testbed using a GUI interface, where the engine can be realistically simulated using industrial-standard driving cycles such as the Worldwide harmonized Light vehicles Test Procedures (WLTP), the New European Driving Cycle (NEDC), the Extra-Urban Driving Cycle (EUDC), and EPA Federal Test Procedure (FTP-75). The engine is modeled using the mean value engine model (MVEM) and is controlled using a proportional-integral (PI)-based boost controller. The GUI interface also allows the user to induce one of the 11 faults of interest, so that their effects on the performance of the engine are better understood. This minimizes the risk of causing permanent damages to the engine and shortening its lifespan, should the tests be conducted onto the actual physical system. This simulation testbed will serve as an excellent platform where researchers can generate critical data to develop and compare current and future research methods for fault diagnosis of automotive engine systems.

\newpage
\processdelayedfloats 
\clearpage

\section[Vehicle Parameters]{Sidebar: Vehicle Parameters}
Tables \ref{VehiclePara} and \ref{GearShifts} show the key vehicle parameters, as well as the estimated gear shift points of the gearbox used to generate the reference inputs for the engine speed $\omega_{eREF}$ and the engine torque $Tq_{eREF}$ in ``\nameref{Controller}.''
\label{AppA}
\begin{table}[t!]
\centering
\caption{\label{VehiclePara}Key vehicle parameters.}
\begin{tabularx}{0.75\textwidth}{p{0.55\textwidth}ll}
\hline
Description & Value & Unit \\ \hline
General vehicle parameters:	& 			& 		\\
~~~Mass of vehicle, $m_v$	& 1700 		& kg 	\\
~~~Drag coefficient, $c_d$	& 0.29		& [–]	\\
~~~Roll coefficient, $c_r$	& 0.013		& [–]	\\
~~~Frontal area, $A_f$		& 2.28		& m$^2$	\\
~~~Wheel radius, $r_w$ 		& 0.3234 	& m 	\\
~~~(assuming tires rated 215/50R17) 	& & 	\\
 & & \\
Gear ratios: 	& 		& 		\\
~~~1st			& 5.250	& [–]	\\
~~~2nd			& 3.029	& [–]	\\
~~~3rd			& 1.950	& [–]	\\
~~~4th			& 1.457	& [–]	\\
~~~5th			& 1.221	& [–]	\\
~~~6th			& 1.000	& [–]	\\
~~~7th			& 0.809	& [–]	\\
~~~8th			& 0.673	& [–]	\\
~~~Reverse		& 4.015	& [–]	\\
~~~Final Drive	& 2.774	& [–]	\\
~~~*Speed per 1000 rpm in 8th gear is 62.9 km/h & & \\ \hline
\end{tabularx}
\end{table}

\begin{table}[t!]
\centering
\caption{\label{GearShifts}Estimated gear shift points of the gearbox.}
\begin{tabularx}{0.7\textwidth}{p{0.1\textwidth}XX} \hline
Gear & Vehicle Speed [km/h] per 1000 rpm & Shifting Engine Speed [rpm] \\ \hline
1st			& 8.070	& 2800	\\
2nd			& 14.00	& 2700	\\
3rd			& 21.70	& 2600	\\
4th			& 29.00	& 2400	\\
5th			& 34.70	& 2200	\\
6th			& 42.30	& 2000	\\
7th			& 52.34	& 1800	\\
8th			& 62.90	& 1600	\\ \hline
\end{tabularx}
\end{table}

\newpage
\processdelayedfloats 
\clearpage

\section[Differential Equations of The Turbocharged Spark Ignited Engine System and Corresponding Engine Parameters]{Sidebar: Differential Equations of The TCSI Engine System and Corresponding Engine Parameters}
\label{AppB}
The testbed is modeled based on the actual engine test bench in the lab, as shown in Figure \ref{Bench}. The total model equations of the TCSI engine are stated below. The first 47 equations are the Turbocharged Spark Ignited (TCSI) engine, as derived in \cite{Lars, EriNie:2014}. The remaining 15 equations ($e_{48} \dots e_{62}$) describe the considered sensors and actuators, as shown in Table \ref{ActSen}. The faults introduced in Table \ref{tab:fault} can be found in equations $e_{15}, e_{17}, e_{23}, e_{25}, e_{29}, e_{50}, e_{56}, e_{57}, e_{59},$ and $e_{60}$, respectively. Table \ref{EngineVar} shows the variables in the engine model, and Table \ref{EnginePara} shows the corresponding key engine parameters.
{
\allowdisplaybreaks
\begin{eqnarray}
&& e_{1}: \dot T_{af} = \frac{R_{a}T_{af}}{p_{af}V_{af}c_{vi}} ((R_{a} + c_{vi})W_{af}T_{af,in} - (R_{a} + c_{vi})W_{c}T_{af} - (W_{af} - W_{c})c_{vi}T_{af}), \nonumber \\
&& e_{2}: \dot p_{af} = \frac{R_{a}T_{af}}{V_{af}}(W_{af} - W_{c}) + \frac{p_{af}}{T_{af}}\dot T_{af}, \nonumber \\
&& e_{3}: \dot T_{c} = \frac{R_{a}T_{c}}{p_{c}V_{ic}c_{vi}} ((R_{a} + c_{vi})W_{c}T_{c,in} - (R_{a} + c_{vi})W_{ic}T_{c} - (W_{c} - W_{ic})c_{vi}T_{c}), \nonumber \\
&& e_{4}: \dot p_{c} = \frac{R_{a}T_{c}}{V_{ic}}(W_{c} - W_{ic}) + \frac{p_{c}}{T_{c}}\dot T_{c}, \nonumber \\
&& e_{5}: \dot T_{ic} = \frac{R_{a}T_{ic}}{p_{ic}V_{ic}c_{vi}} ((R_{a} + c_{vi})W_{ic}T_{ic,in} - (R_{a} + c_{vi})W_{th}T_{ic} - (W_{ic} - W_{th})c_{vi}T_{ic}), \nonumber \\
&& e_{6}: \dot p_{ic} = \frac{R_{a}T_{ic}}{V_{ic}}(W_{ic} - W_{th}) + \frac{p_{ic}}{T_{ic}}\dot T_{ic}, \nonumber \\
&& e_{7}: \dot T_{im} = \frac{R_{a}T_{im}}{p_{im}V_{im}c_{vi}} ((R_{a} + c_{vi})W_{th}T_{im,in} - (R_{a} + c_{vi})W_{ei}T_{im} - (W_{th} - W_{ei})c_{vi}T_{im}), \nonumber \\
&& e_{8}: \dot p_{im} = \frac{R_{a}T_{im}}{V_{im}}(W_{th} - W_{ei}) + \frac{p_{im}}{T_{im}}\dot T_{im}, \nonumber \\
&& e_{9}: \dot T_{em} = \frac{R_{em}T_{em}}{p_{em}V_{em}c_{ve}} ((R_{em} + c_{ve})W_{turbo}T_{em} - (R_{em} + c_{ve})(-W_{eo})T_{t,in} - (W_{turbo} - (-W_{eo}))c_{ve}T_{em}), \nonumber \\
&& e_{10}: \dot p_{em} = \frac{R_{em}T_{em}}{V_{em}}(W_{turbo} - (-W_{eo})) + \frac{p_{em}}{T_{em}}\dot T_{em}, \nonumber \\
&& e_{11}: \dot T_{t} = \frac{R_{em}T_{t}}{p_{t}V_{ex}c_{ve}} ((R_{em} + c_{ve})W_{exh}T_{exh} - (R_{em} + c_{ve})W_{turbo}T_{turbo} - (W_{exh} - W_{turbo})c_{ve}T_{t}), \nonumber \\
&& e_{12}: \dot p_{t} = \frac{R_{em}T_{t}}{V_{ex}}(W_{exh} - W_{turbo}) + \frac{p_{t}}{T_{t}}\dot T_{t}, \nonumber \\
&& e_{13}: \dot \omega_{t} = \frac{1}{J_{t}}((Tq_{t} - Tq_{c}) - \omega_{f}\omega_{t}), \nonumber \\
&& e_{14}: T_{af,in} = \left\{ \begin{array}{ll} T_{amb}, & \mbox{if~~} p_{amb} > p_{af} \\ T_{af}, & \mbox{if~~} p_{af} > p_{amb} \end{array} \right. , \nonumber \\
&& e_{15}: W_{af} = \sqrt{\frac{max(p_{amb},p_{af})}{H_{af}T_{af,in}}} \sqrt{max(p_{amb},p_{af}) - min(p_{amb},p_{af})} + f_{Waf} + f_{paf}, \nonumber \\
&& e_{16}: \Pi_{c} = \frac{p_{c}}{p_{af}}, \nonumber \\
&& e_{17}: W_{c} = \frac{\sqrt{1 - \frac{\Psi_{c}}{\Psi_{cMAX}^2}}\Phi_{cMAX} R_{c}^{3}\omega_{t}p_{af}}{2\pi T_{af} R_{a}} + f_{Wc}, \nonumber \\
&& e_{18}: \Psi_{c} = \frac{4\pi^{2}(R_{a} + c_{vi})T_{af}}{R_{c}^2\omega_{t}^{2}}\left(\Pi_{c}^{\frac{\kappa_{ic} - 1}{\kappa_{ic}}} - 1\right), \nonumber \\
&& e_{19}: \Phi_{c} = \frac{2\pi W_{c}R_{a}T_{af}}{R_{c}^3 \omega_{t}p_{af}}, \nonumber \\
&& e_{20}: \eta_{c} = \frac{\Phi_{c} \eta_{cMAX}}{\Phi_{cMAX}^2} (2\Phi_{cMAX} - \Phi_{c}), \nonumber \\
&& e_{21}: Tq_{c} = \frac{W_{c}(R_{a} + c_{vi})T_{af}}{\eta_{c} \omega_{t}}\left(\Pi_{c}^{\frac{\kappa_{ic} - 1}{\kappa_{ic}}} - 1\right), \nonumber \\
&& e_{22}: T_{ic,in} = \left\{ \begin{array}{ll} T_{c}, & \mbox{if~~} p_{c} > p_{ic} \\ T_{ic}, & \mbox{if~~} p_{ic} > p_{c} \end{array} \right.  ,\nonumber \\
&& e_{23}: W_{ic} = \sqrt{\frac{max(p_{ic},p_{c})}{H_{ic}T_{ic,in}}} \sqrt{max(p_{ic},p_{c}) - min(p_{ic},p_{c})} + f_{Wic}, \nonumber \\
&& e_{24}: T_{th} = \left\{ \begin{array}{ll} T_{ic}, & \mbox{if~~} p_{ic} > p_{im} \\ T_{im}, & \mbox{if~~} p_{im} > p_{ic} \end{array} \right. , \nonumber \\
&& e_{25}: W_{th} = \frac{p_{ic}A_{th}}{\sqrt{T_{ic}R_{a}}}\Psi_{th}(\Pi_{th}) + f_{Wth}, \nonumber \\
&& e_{26}: \Pi_{th} = \frac{p_{im}}{p_{ic}}, \nonumber \\
&& e_{27}: \Pi_{thCRIT} = \left(\frac{2}{\kappa_{th} + 1}\right)^{\frac{\kappa_{th}}{\kappa_{th} - 1}}, \nonumber \\
&& e_{28}: \Psi_{th} = \left\{ \begin{array}{ll} \sqrt{\kappa_{th}}\left(\frac{2}{\kappa_{th} + 1}\right)^{\frac{\kappa_{th} + 1}{2\left(\kappa_{th} - 1\right)}}, & \mbox{if~~} \Pi_{th} \leq \Pi_{thCRIT} \\ \sqrt{\frac{2\kappa_{th}}{\kappa_{th}-1}\left(\Pi_{th}^{\frac{2}{\kappa_{th}}} - \Pi_{th}^{\frac{\kappa_{th}+1}{\kappa_{th}}} \right)}, & \mbox{if~~} \Pi_{th} > \Pi_{thCRIT} \end{array} \right. , \nonumber \\
&& e_{29}: W_{ei} = C_{\eta_{vol}}\frac{r_{c} - \left( \frac{p_{em}}{p_{im}} \right)^{\frac{1}{\kappa_{ei}}}}{r_{c}-1} \frac{V_{im}\omega_{e}p_{im}}{4\pi R_{a}T_{im}} + f_{C_{vol}}, \nonumber \\
&& e_{30}: W_{f} = \frac{W_{ei}}{(A/F)_{s} \lambda}, \nonumber \\
&& e_{31}: W_{eo} = W_{ei} + W_{f}, \nonumber \\
&& e_{32}: T_{eo} = (W_{eo}C_{eo}) + T_{0}, \nonumber \\
&& e_{33}: T_{t,in} = T_{amb} + (T_{eo} - T_{amb})e^{- \frac{h_{tot}\pi d_{pipe} l_{pipe} n_{pipe}}{W_{eo}\frac{R_{em}\kappa_{em}}{\kappa_{em}-1}}}, \nonumber \\
&& e_{34}: T_{wg} = \left\{ \begin{array}{ll} T_{em}, & \mbox{if~~} p_{em} > p_{t} \\ T_{t}, & \mbox{if~~} p_{t} > p_{em} \end{array} \right. , \nonumber \\
&& e_{35}: W_{wg} = \frac{p_{em}u_{wg}c_{D,wg}}{\sqrt{T_{em}R_{em}}}\Psi_{t}(\Pi_{t}), \nonumber \\
&& e_{36}: \Pi_{t} = \frac{p_{t}}{p_{em}}, \nonumber \\
&& e_{37}: \Pi_{tCRIT} = \left(\frac{2}{\kappa_{em} + 1}\right)^{\frac{\kappa_{em}}{\kappa_{em} - 1}}, \nonumber \\
&& e_{38}: \Psi_{t} = \left\{ \begin{array}{ll} \sqrt{\kappa_{em}}\left(\frac{2}{\kappa_{em} + 1}\right)^{\frac{\kappa_{em} + 1}{2\left(\kappa_{em} - 1\right)}}, & \mbox{if~~} \Pi_{t} \leq \Pi_{tCRIT} \\ \sqrt{\frac{2\kappa_{em}}{\kappa_{em}-1}\left(\Pi_{t}^{\frac{2}{\kappa_{em}}} - \Pi_{t}^{\frac{\kappa_{em}+1}{\kappa_{em}}} \right)}, & \mbox{if~~} \Pi_{t} > \Pi_{tCRIT} \end{array} \right. , \nonumber \\
&& e_{39}: BSR = \frac{d_{t}\omega_{t}}{2\sqrt{2c_{p,eg}T_{em}\left(1 - \Pi_{t}^{\frac{1 - \kappa_{em}}{\kappa_{em}}}\right)}}, \nonumber \\
&& e_{40}: \eta_{t} = \eta_{tMAX}\left( 1 - \left( \frac{BSR - BSR_{effMAX}}{BSR_{effMAX}} \right)^{2} \right),  \nonumber \\
&& e_{41}: T_{t,out} = T_{em}\left(1 - \Pi_{t}^{\frac{\kappa_{em}-1}{\kappa_{em}}} \right)\eta_{t}, \nonumber \\
&& e_{42}: W_{t} = \frac{k_{1,t}p_{em}}{\sqrt{T_{em}}}\sqrt{1 - \Pi_{t}^{k_{2,t}}}, \nonumber \\
&& e_{43}: Tq_{t} = \frac{W_{t}c_{p,eg}T_{t,out}}{\omega_{t}}, \nonumber \\
&& e_{44}: W_{turbo} = -(W_{t} + W_{wg}), \nonumber \\
&& e_{45}: T_{turbo} = \frac{W_{t}c_{p,eg}T_{t,out} + W_{wg}c_{p,eg}T_{wg}}{W_{t}c_{p,eg} + W_{wg}c_{p,eg}}, \nonumber \\
&& e_{46}: T_{exh} = \left\{ \begin{array}{ll} T_{amb}, & \mbox{if~~} p_{amb} > p_{t} \\ T_{t}, & \mbox{if~~} p_{t} > p_{amb} \end{array} \right. , \nonumber \\
&& e_{47}: W_{exh} = \sqrt{\frac{max(p_{t},p_{amb})}{H_{ex}T_{exh}}} \sqrt{max(p_{t},p_{amb}) - min(p_{t},p_{amb})}, \nonumber \\
&& e_{48}: u_{p_{amb}} = p_{amb}, \nonumber \\
&& e_{49}: u_{T_{amb}} = T_{amb}, \nonumber \\
&& e_{50}: u_{x_{th}} = A_{th} + f_{x_{th}}, \nonumber \\
&& e_{51}: u_{\omega_{eREF}} = \omega_{eREF}, \nonumber \\
&& e_{52}: u_{x_{wg}} = u_{wg}, \nonumber \\
&& e_{53}: u_{\lambda} = \lambda, \nonumber \\
&& e_{54}: y_{T_{c}} = T_{c}, \nonumber \\
&& e_{55}: y_{p_{c}} = p_{c}, \nonumber \\
&& e_{56}: y_{T_{ic}} = T_{ic} + f_{y_{T_{ic}}}, \nonumber \\
&& e_{57}: y_{p_{ic}} = p_{ic} + f_{y_{p_{ic}}}, \nonumber \\
&& e_{58}: y_{T_{im}} = T_{im}, \nonumber \\
&& e_{59}: y_{p_{im}} = p_{im} + f_{y_{p_{im}}}, \nonumber \\
&& e_{60}: y_{W_{af}} = W_{af} + f_{y_{W_{af}}}, \nonumber \\
&& e_{61}: y_{p_{em}} = p_{em}, \nonumber \\
&& e_{62}: y_{Tq_{e}} = Tq_{e}. \nonumber
\end{eqnarray}
}

\begin{longtable}{lll}
\caption{\label{EngineVar}Variables in the engine model.} \\
\hline
Variable 		& Description 				& Unit \\ \hline
\endhead
\hline
\endfoot
$p_{amb}$		& Ambient pressure							& Pa \\
$p_{af}$		& Air filter pressure						& Pa \\
$p_{c}$			& Compressor pressure						& Pa \\
$p_{ic}$		& Intercooler pressure						& Pa \\
$p_{im}$		& Intake manifold pressure					& Pa \\
$p_{em}$		& Exhaust manifold pressure					& Pa \\
$p_{t}$			& Turbine pressure							& Pa \\
$T_{amb}$		& Ambient temperature						& K \\
$T_{af}$		& Air filter temperature					& K \\
$T_{af,in}$		& Air filter in temperature					& K \\
$T_{c}$			& Compressor temperature					& K \\
$T_{c,in}$		& Compressor in temperature					& K \\
$T_{ic}$		& Intercooler temperature 					& K \\
$T_{ic,in}$		& Intercooler in temperature				& K \\
$T_{im}$		& Intake manifold temperature				& K \\
$T_{im,in}$		& Intake manifold in temperature			& K \\
$T_{em}$		& Exhaust manifold temperature				& K \\
$T_{t}$			& Turbine temperature						& K \\
$T_{t,in}$		& Turbine in temperature					& K \\
$T_{wg}$		& Wastegate temperature						& K \\
$T_{t,out}$		& Temperature difference over turbine		& K \\
$T_{eo}$		& Engine out temperature 					& K \\
$T_{turbo}$		& Turbine and wastegate mixture temperature	& K \\
$T_{exh}$		& Exhaust temperature						& K \\
$W_{af}$		& Mass flow through air filter				& kg/s \\
$W_{c}$			& Mass flow through compressor				& kg/s \\
$W_{ic}$		& Mass flow through intercooler				& kg/s \\
$W_{th}$		& Mass flow through throttle				& kg/s \\
$W_{ei}$		& Mass flow into engine						& kg/s \\
$W_{f}$			& Fuel mass flow							& kg/s \\
$W_{eo}$		& Mass flow out from engine					& kg/s \\
$W_{wg}$		& Mass flow through wastegate				& kg/s \\
$W_{turbo}$		& Mass flow of turbine and wastegate mixture& kg/s \\
$W_{exh}$		& Mass flow through exhaust					& kg/s \\
$\omega_{t}$	& Turbine speed								& rad/s \\
$\omega_{e}$	& Engine speed								& rad/s \\
$Tq_{t}$		& Turbine torque							& N$\cdot$m \\
$Tq_{c}$		& Compressor torque							& N$\cdot$m \\
$\Pi_{c}$		& Pressure ratio in compressor 				& [--] \\
$\Pi_{th}$		& Pressure ratio in throttle 				& [--] \\
$\Pi_{thCRIT}$	& Critical pressure ratio in throttle		& [--] \\
$\Pi_{t}$		& Pressure ratio in turbine 				& [--] \\
$\Pi_{tCRIT}$	& Critical pressure ratio in turbine 		& [--] \\
$\Phi_{c}$		& Energy transfer coefficient				& [--] \\
$\Psi_{c}$		& Compressor flow coefficient 				& \% \\
$\Psi_{th}$		& Throttle flow coefficient 				& \% \\
$\Psi_{t}$		& Turbine flow coefficient  				& \% \\
$\eta_{c}$		& Compressor efficiency						& \% \\
$\eta_{t}$		& Turbine efficiency						& \% \\
BSR				& Blade speed ratio							& [--] \\
\hline
\end{longtable}

\begin{longtable}{lrl}
\caption{\label{EnginePara}Key parameters of the engine in this testbed.} \\
\hline
Description & Value & Unit \\ \hline
\endhead
\hline
\endfoot
Ambient Air Data:										&									&				\\
Ratio of specific heats, $\kappa_{ic}$					& 1.4 								& [–]	\\
Gas constant, $R_{a}$									& 287.2 							& J/(kg$\cdot$K) \vspace{3mm} \\
Engine Block:											& 									&		\\
Bore, $B$												& 0.0831							& m		\\
Displacement volume, $V_{d}$							& 0.0018							& m$^3$	\\
Number of cylinders, $n_{cyl}$							& 4									& [–]	\\
Number of revolutions per power stroke, $n_{r}$			& 2									& [-]	\\
Compression ratio, $r_{c}$								& 9.5								& [–]	\\
Boost layout, $\Pi_{bl}$ 								& 2									& [–]	\\
Factor for auxiliary devices, $\xi_{aux}$				& 1	 								& [–]	\\
Gross efficiency, $\eta_{ig}$				  			& 0.4								& [–]	\\
Stoichiometric factor air to fuel, $(A/F)_{s}$			& 15.1 								& [–]	\\
Air to fuel ratio, $\lambda$							& 1			 						& [–]	\\
Volumetric efficiency constant, $C_{\eta_{vol}}$		& 0.8								& [–]	\\
Ratio of specific heats, $\kappa_{ei}$					& 1.3								& [–]	\\
Gas constant, $R_{em}$									& 290								& J/(kg$\cdot$K)\\
Intake manifold volume, $V_{im}$						& 0.0018							& m$^3$	\\
Exhaust manifold volume, $V_{em}$						& 0.0025							& m$^3$	\\
BMEP parameter 1, $C_{Tq1}$								& $0.2$$\times$$10^6$				& Pa		\\
BMEP parameter 2, $C_{Tq2}$								& $1.2$$\times$$10^6$				& Pa		\\
Temperature at 0 mass flow, $T_{0}$						& 1100								& K		\\
Temperature change with mass flow, $C_{eo}$				& 3000								& K$\cdot$s/kg	\\
Fuel lower heating value, $q_{HV}$						& $4.4$$\times$$10^7$ 				& J/kg  \\
Measurement constant, $a_{0}$							& $1.1647^{-5}$						& [-] \\
Measurement constant, $a_{1}$							& $3.0718^{-5}$						& [-] \\
Measurement constant, $a_{2}$							& $0.0029$							& [-] \vspace{3mm} \\
Air Filter:												& 									&		\\
Volume, $V_{af}$										& 0.01								& m$^3$	\\
Flow resistance, $H_{af}$					  			& $2$$\times$$10^8$ 				& [–] 	\\
Linearization limit, $p_{lin,af}$				 		& 2000 								& Pa \vspace{3mm} \\
Compressor:												& 									&		\\
Volume, $V_{c}$											& 0.005								& m$^3$	\\
Diameter, $D_{c}$	 									& 0.06 								& m 		\\
Maximum efficiency, $\eta_{cMAX}$						& 0.8 								& [–]	\\
Minimum efficiency, $\eta_{cMIN}$						& 0.3 								& [–]	\\
Maximum flow coefficient, $\Phi_{cMAX}$					& 0.12 								& [–]	\\
Head parameter, $\Psi_{cMAX}$							& 2.3 								& [–] \vspace{3mm} \\
Throttle:												&									&		\\
Ratio of specific heats, $\kappa_{th}$					& 2 								& [–]	\\
Maximum pressure ratio, $\Pi_{thMAX}$					& 0.9								& [–]	\vspace{3mm} \\ 
Intercooler:											& 									&		\\
Volume, $V_{ic}$										& 0.005								& m$^3$	\\
Flow resistance, $H_{ic}$								& $4$$\times$$10^8$ 				& [–]	\\
Linearization limit, $p_{lin,ic}$						& 500 								& Pa		\\
Heat transfer coefficient, $h_{ic}$						& 0.8 								& [–] \vspace{3mm} \\
Regulated pressure drop across throttle, $\Delta p_{thREF}$ & 10000						& Pa 	\\
\multicolumn{2}{l}{Exhaust and Turbine Inlet:} 			&		\\
Volume , $V_{ex}$										& 0.02								& m$^3$	\\
Ratio of specific heats, $\kappa_{em}$					& 1.3 								& [–]	\\
Gas constant, $R_{em}$									& 290 								& J/(kg$\cdot$K) \\
Pipe diameter, $d_{pipe}$ 								& 0.04								& m		\\
Pipe length, $l_{pipe}$									& 0.45								& m		\\
Number of parallel pipes, $n_{pipe}$					& 4 								& [–]	\\
External heat transfer coefficient, $h_{ext}$			& 95								& W/(m$^2\cdot$K)\\
Dynamic viscosity, $\mu_{em}$ 							& $4$$\times$$10^{-5}$				& kg/(m$\cdot$s)	\\
Thermal conductivity, $k_{em}$							& 0.07								& W/(m$\cdot$K)	\\
Flow resistance, $H_{ex}$								& $3$$\times$$10^8$					& [–] 	\\
Linearization limit, $p_{lin,ex}$						& 300								& Pa \vspace{3mm} \\
Turbocharger:											& 									&		\\
Friction coefficient, $\omega_{f}$						& $1$$\times$$10^{-6}$				& [–]	\\
Inertia of turbocharger, $J_{t}$						& $3$$\times$$10^{-5}$				& kg$\cdot$m$^2$\\
Initial speed, $\omega_{tINIT}$							& 3000								& rad/s	\\
Minimum speed, $\omega_{tMIN}$							& 2000								& rad/s	\\
Maximum speed, $\omega_{tMAX}$							& $2.4$$\times$$10^4$				& rad/s	\vspace{3mm} \\
\multicolumn{2}{l}{Turbine and Wastegate:}				&		\\
Turbine diameter, $d_{t}$								& 0.05								& m		\\
Specific heat of gas, $c_{p,eg}$						& 1200   							& J/(kg$\cdot$K) \\
Ratio of specific heats, $\kappa_{em}$					& 1.3								& [–]	\\
Maximum turbine efficiency, $\eta_{tMAX}$				& 0.75								& [–]	\\
Minimum turbine efficiency, $\eta_{tMIN}$				& 0.3								& [–]	\\
BSR at maximum turbine efficiency, $BSR_{effMAX}$		& 0.7								& [–]	\\
Mass flow constant 1, $k_{1,t}$							& 0.017								& [–]	\\
Mass flow constant 2, $k_{2,t}$							& 1.4								& [–]	\\
Discharge coefficient, $c_{D,wg}$						& 0.9								& [–]	\\
Maximum wastegate area, $A_{wgMAX}$						& $3.5$$\times$$10^{-4}$			& m$^2$ \vspace{3mm} \\ \hline
\end{longtable}

\newpage
\processdelayedfloats 
\clearpage

\section[Modeling Mass Flow Fault Caused By A Leakage]{Sidebar: Modeling Mass Flow Fault Caused By A Leakage}
\label{sec:Leak}
The intensity of a mass flow fault caused by a leakage is determined by the area of the leakage orifice, which is usually measured in mm$^{2}$. The bigger the leakage area, the higher the mass flow through the leakage and the higher the pressure difference between both sides of the orifice (hence, the higher intensity of the fault). Conventionally, a mass flow fault is physically induced by drilling a hole in the specific component of the engine test bench system. Tests are then performed by running the engine through driving cycles, and the effects of the fault on the performance of the engine are analyzed. The leakage orifice is then sealed using a screw plug to ”disable” the fault and return the engine to its nominal operation mode. Therefore, by simulating the mass flow fault using a simulation testbed in Matlab/Simulink, it removes the need for the engine system to be physically modified to accommodate such fault (which could lead to irreversible damages).

The mathematical modeling of the mass flow fault for compressible flows was briefly discussed in \cite{nyberg2002model}, where the flow through the leakage is described using
\begin{equation}
W_{leak} = k_{leak}\frac{p_{high}}{\sqrt{T_{amb}}}\Psi\left(\frac{p_{low}}{p_{high}}\right),
\end{equation}
where $k_{leak}$ is the area of the leakage orifice (mm$^{2}$), $T_{amb}$ the ambient temperature (K), and $p_{high}$ and $p_{low}$ are the higher and lower pressures (Pa) on either side of the leakage. The function $\Psi\left(\frac{p_{low}}{p_{high}}\right)$ is defined as
\begin{equation}
\begin{array}{l}
\Psi\left(\frac{p_{low}}{p_{high}}\right) =
\left\{\begin{array}{l}
\sqrt{\frac{2\kappa}{\kappa - 1} \left\{ \left( \frac{p_{low}}{p_{high}}\right)^{2/\kappa} - \left(\frac{p_{high}}{p_{low}}\right)^{(\kappa+1)/\kappa}\right\} },
~~\mbox{if } \left(\frac{p_{low}}{p_{high}}\right) \geq \left(\frac{2}{\kappa + 1}\right)^{\kappa/(\kappa -1)} \\ \\
\sqrt{\kappa \left(\frac{2}{\kappa + 1}\right)^{(\kappa + 1)/(\kappa - 1)}}, ~~~\mbox{otherwise}
\end{array}\right. ,
\end{array}
\end{equation}
where $\kappa$ is the specific heat ratio in the affected part of the engine system.

\newpage
\processdelayedfloats 
\clearpage

\section[Structural Model and Fault Isolation Model: A General Tutorial]{Sidebar: Structural Model and FIM: A General Tutorial}
\label{SMFIM}
Let's consider a dc motor system shown in Figure \ref{fig:DCMotor}, which is modeled using the following equations:
\begin{figure}[t!]
\centering
\includegraphics[width=0.6\columnwidth]{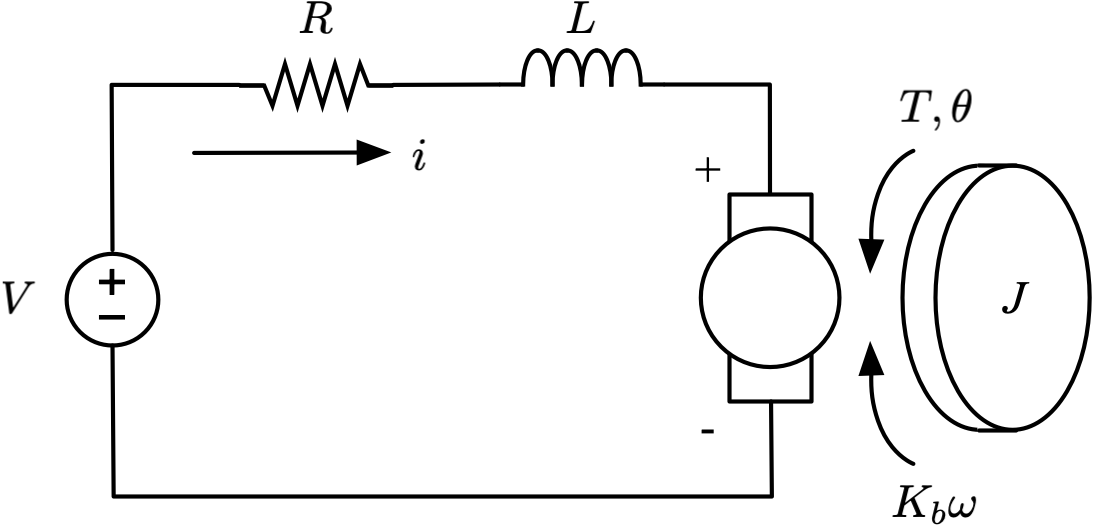}
\caption{A general dc motor system commonly used in control engineering studies.}
\label{fig:DCMotor}
\end{figure}
\begin{eqnarray}
&& e_{1}: V = i(R + f_{R}) + L \frac{di}{dt} + K_{a}i\omega , \label{eq:DCV}\\
&& e_{2}: T_{m} = K_{a}i^{2} , \\
&& e_{3}: J \frac{d\omega}{dt} = \Delta T - K_{b}\omega, \\
&& e_{4}: \Delta T = T_{m} - T_{L}, \\
&& e_{5}: \frac{d\theta}{dt} = \omega, \\
&& e_{6}: \frac{d\omega}{dt} = \alpha, \\
&& e_{7}: y_{i} = i + f_{i}, \label{eq:fyi} \\
&& e_{8}: y_{\omega} = \omega + f_{\omega}, \\
&& e_{9}: y_{\Delta} = \Delta T + f_{\Delta}, \label{eq:fydelta}
\end{eqnarray}
where $V$ is the input voltage, $R$ the resistance, $L$ the inductance, and $i$ the current in the armature circuit. On the mechanical side of the system, $T_{m}$ is the motor torque, $T_{L}$ the torque of the load, $J$ the moment of inertia of the rotor, $K_{a}$ the motor torque constant, and $K_{b}$ the back EMF constant. The rotational displacement of the motor is $\theta$, while $\omega$ and $\alpha$ are the rotational velocity and acceleration, respectively. The states of the system are $\{i, \theta, \omega, \alpha, T_{m}, T_{L}, \Delta T\}$, and the measurable outputs are $\{i, \omega, \Delta T\}$. It is assumed that there is a system fault $f_{R}$ representing inconsistency in the value of the resistance $R$. It is also assumed that all outputs are potentially faulty through $f_{i}, f_{\omega}$, and $f_{\Delta}$, respectively, as shown in (\ref{eq:fyi})--(\ref{eq:fydelta}).

Using structural model, the relationships among the unknown variables (states), known variables (inputs and outputs), and faults in the system can be explained using Table \ref{tab:DCSM}. In the table, an `X' is placed in the corresponding columns where the variables or faults are used to explain each equation in (\ref{eq:DCV})--(\ref{eq:fydelta}). For example, the states $i$ and $\omega$ (as well as the fault $f_{R}$) are used in equation $e_{1}$ in (\ref{eq:DCV}) to express the input voltage $V$.
\begin{table}[t]
\caption{Structural model of the dc motor system as modeled using (\ref{eq:DCV})--(\ref{eq:fydelta}). Here, the relationships among the unknown variables (states), known variables (inputs and outputs), and faults in the system can be explained by placing an `X' in the corresponding columns where the variables or faults are used to explain each equation.}
\begin{center}
\begin{tabular}{l|ccccccc|cccc|cccc}
\hline
& $i$ & $\theta$ & $\omega$ & $\alpha$ & $T_{m}$ & $T_{L}$ & $\Delta T$ & $V$ & $y_{i}$ & $y_{\omega}$ & $y_{\Delta}$ & $f_{R}$ & $f_{i}$ & $f_{\omega}$ & $f_{\Delta}$ \\ \hline
$e_{1}$ & X &   & X &   &   &   &   & X &   &   &   & X &   &   &   \\
$e_{2}$ & X &   &   &   & X &   &   &   &   &   &   &   &   &   &   \\
$e_{3}$ &   &   & X &   &   &   & X &   &   &   &   &   &   &   &   \\
$e_{4}$ &   &   &   &   & X & X & X &   &   &   &   &   &   &   &   \\
$e_{5}$ &   & X & X &   &   &   &   &   &   &   &   &   &   &   &   \\
$e_{6}$ &   &   & X & X &   &   &   &   &   &   &   &   &   &   &   \\
$e_{7}$ & X &   &   &   &   &   &   &   & X &   &   &   & X &   &   \\
$e_{8}$ &   &   & X &   &   &   &   &   &   & X &   &   &   & X &   \\
$e_{9}$ &   &   &   &   &   &   & X &   &   &   & X &   &   &   & X \\ \hline
\end{tabular}
\end{center}
\label{tab:DCSM}
\end{table}

By performing canonical decomposition onto the unknown variables in Table \ref{tab:DCSM}, the structural model can then be remodeled as Table \ref{tab:DCSM2}. The fault isolation matrix (FIM) is then obtained by extracting the bottom-right section of the the structural model in Table \ref{tab:DCSM2}. It can be seen that for the dc motor system, the pair $\{f_{R}, f_{i}\}$ are not isolable from each other. See Table \ref{tab:DCFIM}.
\begin{table}[t]
\caption{A remodel of the structural model in Table \ref{tab:DCSM} by performing canonical decomposition on the unknown variables.}
\begin{center}
\begin{tabular}{l|cccccccc}
\hline
        & $\theta$               & $\alpha$               & $T_{L}$                & $T_{m}$                & $i$                    & $\Delta T$             & $\omega$ \\ \hline
$e_{5}$ & \multicolumn{1}{c|}{X} &                        &                        &                        &                                                &                                                & X                                              \\ \cline{2-3}
$e_{6}$ & \multicolumn{1}{c|}{}  & \multicolumn{1}{c|}{X} &                        &                        &                                                &                                                & X                                              \\ \cline{3-4}
$e_{4}$ &                        & \multicolumn{1}{c|}{}  & \multicolumn{1}{c|}{X} & X                      &                                                & X                                              &                                                \\ \cline{4-5}
$e_{2}$ &                        &                        & \multicolumn{1}{c|}{}  & \multicolumn{1}{c|}{X} & X                                              &                                                &                                                \\ \cline{5-8}
$e_{1}$ &                        &                        &  					   \multicolumn{2}{r|}{$f_{R} \longrightarrow$}  	& \multicolumn{1}{c|}{\cellcolor[HTML]{C0C0C0}X} 	& \multicolumn{1}{c|}{}                          & \multicolumn{1}{c|}{X}                         \\
$e_{7}$ &                        &                        &                        \multicolumn{2}{r|}{$f_{i} \longrightarrow$}  	& \multicolumn{1}{c|}{\cellcolor[HTML]{C0C0C0}X} 	& \multicolumn{1}{c|}{}                          & \multicolumn{1}{c|}{}                          \\ \cline{6-7}
$e_{3}$ &                        &                        &                        & \multicolumn{1}{c|}{}  & \multicolumn{1}{c|}{}                          & \multicolumn{1}{c|}{\cellcolor[HTML]{C0C0C0}X} & \multicolumn{1}{c|}{X}                         \\
$e_{9}$ &                        &                        &                        \multicolumn{2}{r|}{$f_{\Delta} \longrightarrow$}  & \multicolumn{1}{c|}{}                          	& \multicolumn{1}{c|}{\cellcolor[HTML]{C0C0C0}X} & \multicolumn{1}{c|}{}                          \\ \cline{6-8}
$e_{8}$ &                        &                        &                        \multicolumn{2}{r|}{$f_{\omega} \longrightarrow$}  &                                                	& \multicolumn{1}{c|}{}                          & \multicolumn{1}{c|}{\cellcolor[HTML]{C0C0C0}X} \\ \hline
\end{tabular}
\end{center}
\label{tab:DCSM2}
\end{table}

\begin{table}[t]
\caption{The fault isolation matrix (FIM) of the dc motor system in (\ref{eq:DCV})--(\ref{eq:fydelta}). It can be seen that the pair $\{f_{R}, f_{i}\}$ are not isolable from each other.}
\begin{center}
\begin{tabular}{l|cccc} \hline
& $f_{R}$ & $f_{i}$ & $f_{\Delta}$ & $f_{\omega}$ \\ \hline
$f_{R}$ 		& X & X &   &   \\
$f_{i}$ 		& X & X &   &   \\
$f_{\Delta}$ 	& 	& 	& X &  	\\
$f_{\omega}$ 	& 	& 	& 	& X \\ \hline
\end{tabular}
\end{center}
\label{tab:DCFIM}
\end{table}

\newpage
\processdelayedfloats 
\clearpage

\section{Author Biography}
Kok Yew Ng (mark.ng@ulster.ac.uk) is currently a lecturer in the School of Engineering, Ulster University, UK, where he also conducts research at the Nanotechnology and Integrated Bioengineering Centre (NIBEC). He received the B.Eng degree (Hons) in electrical \& computer systems engineering and the Ph.D. degree from Monash University in 2006 and 2009, respectively. From 2014 to 2015, he was a postdoctoral researcher with the Division of Vehicular Systems, Link\"oping University, Sweden, where he focused on the design and development of advanced fault diagnosis schemes using model-based and data-driven methods on vehicular engines with Volvo Cars. He is also an adjunct senior research fellow in the School of Engineering, Monash University Malaysia. His research interests are fault diagnosis, sliding-mode observer, vehicular systems, and data analytics for anomaly detection and classification.

Erik Frisk received the Ph.D. degree from Link\"oping University, Link\"oping, Sweden, in 2001. He is currently a professor in the Department of Electrical Engineering, Link\"oping University. His main research interests are fault diagnosis, fault isolation, and prognostics using model-based techniques and data-driven approaches. Control of autonomous vehicles is also an active research area.

Mattias Krysander is an associate professor in the Department of Electrical Engineering, Link\"oping University, Sweden. His research interests include model-based and data-driven diagnosis and prognosis. As a way to cope with the complexity and size of industrial systems (mainly vehicle systems), he has used structural representations of models and developed graph theoretical methods for assisting design of diagnosis systems and for fault isolation and sensor placement analysis.

Lars Eriksson is a full professor of vehicular systems at Link\"oping University. He received the M.Sc. degree in electrical engineering in 1995 and the Ph.D. degree in vehicular systems in 1999, both from Link\"oping University. From 2000 to 2001, he was a postdoctoral researcher in the Measurement and Control group at Swiss Federal Institute of Technology (ETH) in Zurich. Since then, he has been a faculty member in the Department of Vehicular Systems. He is currently managing the engine laboratory at Vehicular Systems. His research interests are modeling, simulation, and control of internal combustion engines for vehicle propulsion in general, with a focus on downsizing and supercharging concepts for improved fuel economy. His contributions are foremost on engine control and control-oriented modeling of combustion engines, where he was the first to demonstrate real-time control of the combustion timing using information obtained from the ion current. He is also active in the academic societies and is currently a member of the IFAC Technical Board as chair for the Coordinating Committee CC 7 on Transportation and Vehicle Systems. He is also associate editor for the Control Engineering Practice, and has served as adjoint technical editor for several conferences, including the IFAC World Congresses, Advances in Automotive Control (AAC), and E-CoSM.

\end{document}

%% file: Figures/Figure3.tex
\tikzstyle{block} = [draw, fill = blue!20, rectangle, minimum height = 12em, minimum width = 5em]
\tikzstyle{block2} = [draw, fill = green!20, rectangle, minimum height = 4em, minimum width = 5em]
\tikzstyle{block3} = [draw, fill = pink!60, rectangle, minimum height = 6em, minimum width = 5em]
\tikzstyle{block4} = [draw, fill = yellow!30, rectangle, minimum height = 3em, minimum width = 5em]
\tikzstyle{sum} = [draw, fill = gray!50, circle]
\tikzstyle{input} = [coordinate]
\tikzstyle{output} = [coordinate]
\tikzstyle{pinstyle} = [pin edge={to-,thick,black}] 
\tikzstyle{ann} = [above, text width=5em]

\def\blockdist{2.3}
\def\edgedist{2.5}

\hspace{-1.5cm}
\begin{tikzpicture}[auto, node distance=2.4cm, >=latex']

    \node [block3] (controller) {\begin{tabular}{c} Boost \\ Controller \end{tabular}};
    \node [block, right of = controller, node distance = 6cm] (engine) {Engine}; 
    \node [block2, below of = engine, node distance = 4cm] (mux) {Mux};
    
    \path (controller.150)+(-\blockdist,0) node (weref) [input, node distance = 0.5cm] {};
    \path (controller.-180)+(-\blockdist,0) node (Tqref) [input] {};   
    \path (engine.-145)+(-\blockdist,0) node (lambda) [input] {};   
    \path (engine.-125)+(-\blockdist,0) node (Tamb) [input] {};        
    \path (engine.-115)+(-\blockdist,0) node (pamb) [input] {};       

    \draw [->] (weref) -- node [pos = 0.2] {$\omega_{eREF}$} node [name = werefOut, pos = 0.6] {} (controller.west |- weref);   		
    \draw [->] (Tqref) -- node [pos = 0.2] {$Tq_{eREF}$} (controller.west |- Tqref);  
    \draw [->] (lambda) -- node[pos = 0.1] {$\lambda$} (engine.-145 |- lambda);
    \draw [->] (Tamb) -- node[pos = 0.2] {$T_{amb}$} (engine.-125 |- Tamb);    
    \draw [->] (pamb) -- node[pos = 0.2] {$p_{amb}$} (engine.-115 |- pamb);    
    \draw [->] (controller.30) -- node [pos = 0.2] {$A_{th}$} (engine.west |- controller.30);   	
    \draw [->] (controller.0) -- node [pos = 0.2] {$u_{wg}$} (engine.west |- controller.0);  
    \draw [->] (weref)+(1.5,0) |- (engine.120); 
    \draw [->] (engine.0) -- node [ann] {} + (\edgedist,0) node [right] (Tim) {$T_{im}$};	 
    \draw [->] (engine.65) -- node [ann] {} + (\edgedist,0) node [right] {$T_{c}$};	    
    \draw [->] (engine.58) -- node [ann] {} + (\edgedist,0) node [right] {$p_{c}$};	 
    \draw [->] (engine.47) -- node [ann] {} + (\edgedist,0) node [right] {$T_{ic}$};	    
    \draw [->] (engine.28) -- node [ann] {} + (\edgedist,0) node [right] (pic) {$p_{ic}$};	    
    \draw [->] (engine.-65) -- node [ann] {} + (\edgedist,0) node [right] (pem) {$p_{em}$};	    
    \draw [->] (engine.-58) -- node [ann] {} + (\edgedist,0) node [right] (Tqe) {$Tq_{e}$};	    
    \draw [->] (engine.-47) -- node [ann] {} + (\edgedist,0) node [right] {$W_{af}$};	    
    \draw [->] (engine.-28) -- node [ann] {} + (\edgedist,0) node [right] (pim) {$p_{im}$};	    

	\draw (mux) -| (-2,-0.7);
	\draw [->] (-2,-0.7) -- (-1.2,-0.7);
	\draw [->] (pim)+(-1.7,0) |- (mux);
	\draw [->] (pic)+(-2.4,0) |- (mux.30);
	\draw [->] (pem)+(-1,0) |- (mux.-30);
	\draw [->] (Tim)+(-2.1,0) |- (mux.15);
	\draw [->] (Tqe)+(-1.35,0) |- (mux.-15);
\end{tikzpicture}

%% file: Figures/Figure7.tex
\tikzstyle{block} = [draw, fill = blue!20, rectangle, minimum height = 3em, minimum width = 5em]
\tikzstyle{block2} = [draw, fill = green!20, rectangle, minimum height = 3em, minimum width = 5em]
\tikzstyle{block3} = [draw, fill = pink!60, rectangle, minimum height = 3em, minimum width = 5em]
\tikzstyle{block4} = [draw, fill = yellow!30, rectangle, minimum height = 3em, minimum width = 5em]
\tikzstyle{sum} = [draw, fill = gray!50, circle]
\tikzstyle{input} = [coordinate]
\tikzstyle{output} = [coordinate]
\tikzstyle{pinstyle} = [pin edge={to-,thick,black}] 

\hspace{-1cm}
\begin{tikzpicture}[auto, node distance=2.4cm, >=latex']
	\node [block4] (ref) {Reference};
    \node [sum, right of = ref] (sum1) {};
    \node [block3, right of = sum1] (controller) {\begin{tabular}{c} Boost \\ Controller \end{tabular}};
    \node [sum, right of = controller] (sum2) {};
    \node [block, right of = sum2] (engine) {Engine}; 
    \node [sum, right of = engine] (sum3) {};
    
    \node [input, above of = sum2, node distance = 1.5cm] (fu) {};
    \node [input, above of = sum3, node distance = 1.5cm] (fy) {};
    \node [input, above of = engine, node distance = 1.5cm] (fx) {};
    
    \node [block2, below of = engine] (observer) {Observer};
    \node [sum, right of = observer, node distance = 3.5cm] (sum4) {};
    \node [output, right of = sum4, node distance = 1cm] (output) {};

    \draw [->] (ref) -- node [pos=0.8] {$+$} (sum1);     
    \draw [->] (controller) -- node [pos = 0.85] {$+$} node [pos = 0.3] {$u$} node [name = u, pos = 0.5] {} (sum2);
    \draw [->] (fu) -- node [above, pos = 0] {\begin{tabular}{c} Actuator \\ fault \end{tabular}} node[pos = 0.8] {$+$} (sum2);
	\draw [->] (fy) -- node [above, pos = 0] {\begin{tabular}{c} Sensor \\ fault \end{tabular}} node[pos = 0.8] {$+$} (sum3);
	\draw [->] (fx) -- node [above, pos = 0] {\begin{tabular}{c} Variable \\ fault \end{tabular}} (engine);
    \draw [->] (sum1) -- node{} (controller);
    \draw [->] (engine) -- node [pos = 0.85] {$+$} (sum3);
    \draw [->] (sum3) -| node {$y$} node [pos = 0.95] {$-$} (sum4);    
    \draw [->] (u) |- (observer);
    \draw [->] (sum2) -- node {} (engine);    
    \draw [->] (observer) -- node {$\hat y$} node [pos = 0.9] {$+$} (sum4);    
    \draw [->] (sum4) -- node [name = r, pos = 0.99] {$r$} (output);
    \draw (sum3) -| node {} (13.1,3);
    \draw [->] (13.1,3) -| node [pos = 0.95] {$-$} (sum1);
    
    \draw [red, dashed, very thick] (7.7,-3.4) -- (13.8,-3.4) -- (13.8,-1.4) -- (7.7,-1.4) -- (7.7,-3.4);
    \node at (9.2,-3.8) [above=5mm, right=0mm] {Residuals Generator};
    \draw [blue, dotted, very thick] (-1.5,-1) -- (13.8,-1) -- (13.8,3.5) -- (-1.5,3.5) -- (-1.5,-1);    
    \node at (3.5,3.8) [above=5mm, right=0mm] {Closed-loop Engine Control System};    
\end{tikzpicture}

%% file: main.bbl
\begin{thebibliography}{10}
\providecommand{\url}[1]{#1}
\csname url@samestyle\endcsname
\providecommand{\newblock}{\relax}
\providecommand{\bibinfo}[2]{#2}
\providecommand{\BIBentrySTDinterwordspacing}{\spaceskip=0pt\relax}
\providecommand{\BIBentryALTinterwordstretchfactor}{4}
\providecommand{\BIBentryALTinterwordspacing}{\spaceskip=\fontdimen2\font plus
\BIBentryALTinterwordstretchfactor\fontdimen3\font minus
  \fontdimen4\font\relax}
\providecommand{\BIBforeignlanguage}[2]{{%
\expandafter\ifx\csname l@#1\endcsname\relax
\typeout{** WARNING: IEEEtran.bst: No hyphenation pattern has been}%
\typeout{** loaded for the language `#1'. Using the pattern for}%
\typeout{** the default language instead.}%
\else
\language=\csname l@#1\endcsname
\fi
#2}}
\providecommand{\BIBdecl}{\relax}
\BIBdecl

\bibitem{gerler1995model}
J.~{Gerler}, M.~{Costin}, {Xiaowen Fang}, Z.~{Kowalczuk}, M.~{Kunwer}, and
  R.~{Monajemy}, ``Model based diagnosis for automotive engines-algorithm
  development and testing on a production vehicle,'' \emph{IEEE Transactions on
  Control Systems Technology}, vol.~3, no.~1, pp. 61--69, March 1995.

\bibitem{kher2001automobile}
S.~{Kher}, P.~K. {Chande}, and P.~C. {Sharma}, ``Automobile engine fault
  diagnosis using neural network,'' in \emph{ITSC 2001. 2001 IEEE Intelligent
  Transportation Systems. Proceedings (Cat. No.01TH8585)}, Oakland, CA, USA,
  Aug. 2001, pp. 492--495.

\bibitem{nyberg2002model}
M.~{Nyberg}, ``Model-based diagnosis of an automotive engine using several
  types of fault models,'' \emph{IEEE Transactions on Control Systems
  Technology}, vol.~10, no.~5, pp. 679--689, Sep. 2002.

\bibitem{murphey2003automotive}
Y.~L. {Murphey}, J.~A. {Crossman}, {ZhiHang Chen}, and J.~{Cardillo},
  ``{Automotive fault diagnosis - part II: a distributed agent diagnostic
  system},'' \emph{IEEE Transactions on Vehicular Technology}, vol.~52, no.~4,
  pp. 1076--1098, July 2003.

\bibitem{denton2016advanced}
T.~Denton, \emph{{Advanced Automotive Fault Diagnosis}}.\hskip 1em plus 0.5em
  minus 0.4em\relax London: Routledge, 2017.

\bibitem{goodloe2010monitoring}
A.~E. Goodloe and L.~Pike, ``Monitoring distributed real-time systems: A survey
  and future directions,'' \emph{NASA Technical Reports Server}, 2010.

\bibitem{scacchioli2006model}
A.~Scacchioli, G.~Rizzoni, and P.~Pisu, ``{Model-based fault detection and
  isolation in automotive electrical systems},'' in \emph{Proceedings of the
  ASME 2006 International Mechanical Engineering Congress and Exposition.
  Dynamic Systems and Control, Parts A and B.}\hskip 1em plus 0.5em minus
  0.4em\relax Chicago, Illinois, USA: ASME, Nov. 2006, pp. 315--324.

\bibitem{weber1999multiple}
P.~Weber, S.~Gentil, P.~Ripoll, and L.~Foulloy, ``{Multiple fault detection and
  isolation},'' in \emph{{14th IFAC World Congress}}, Beijing, China, 1999, pp.
  223--228.

\bibitem{da2012knowledge}
J.~C. da~Silva, A.~Saxena, E.~Balaban, and K.~Goebel, ``A knowledge-based
  system approach for sensor fault modeling, detection and mitigation,''
  \emph{Expert Systems with Applications}, vol.~39, no.~12, pp.
  10\,977--10\,989, 2012.

\bibitem{tang2008prognostics}
L.~{Tang}, G.~J. {Kacprzynski}, K.~{Goebel}, A.~{Saxena}, B.~{Saha}, and
  G.~{Vachtsevanos}, ``Prognostics-enhanced automated contingency management
  for advanced autonomous systems,'' in \emph{2008 International Conference on
  Prognostics and Health Management}, Denver, CO, USA, Oct. 2008, pp. 1--9.

\bibitem{loureiro2012bond}
R.~{Loureiro}, R.~{Merzouki}, and B.~{Ould-Bouamama}, ``Bond graph model based
  on structural diagnosability and recoverability analysis: Application to
  intelligent autonomous vehicles,'' \emph{IEEE Transactions on Vehicular
  Technology}, vol.~61, no.~3, pp. 986--997, March 2012.

\bibitem{Loureiro2014integration}
R.~{Loureiro}, S.~{Benmoussa}, Y.~{Touati}, R.~{Merzouki}, and
  B.~{Ould-Bouamama}, ``{Integration of fault diagnosis and fault-tolerant
  control for health monitoring of a class of MIMO intelligent autonomous
  vehicles},'' \emph{IEEE Transactions on Vehicular Technology}, vol.~63,
  no.~1, pp. 30--39, Jan. 2014.

\bibitem{isermann1999hardware}
R.~Isermann, J.~Schaffnit, and S.~Sinsel, ``Hardware-in-the-loop simulation for
  the design and testing of engine-control systems,'' \emph{Control Engineering
  Practice}, vol.~7, no.~5, pp. 643--653, 1999.

\bibitem{butler1999matlab}
K.~L. {Butler}, M.~{Ehsani}, and P.~{Kamath}, ``{A Matlab-based modeling and
  simulation package for electric and hybrid electric vehicle design},''
  \emph{IEEE Transactions on Vehicular Technology}, vol.~48, no.~6, pp.
  1770--1778, Nov. 1999.

\bibitem{assanis2000validation}
D.~Assanis, Z.~Filipi, S.~Gravante, D.~Grohnke, X.~Gui, L.~Louca, G.~Rideout,
  J.~Stein, and Y.~Wang, ``{Validation and use of Simulink integrated, high
  fidelity, engine-in-vehicle simulation of the International Class VI
  truck},'' \emph{SAE Transactions}, vol. 109, pp. 384--399, 2000.

\bibitem{lee2003cost}
W.~Lee, M.~Yoon, and M.~Sunwoo, ``A cost- and time-effective
  hardware-in-the-loop simulation platform for automotive engine control
  systems,'' \emph{Proceedings of the Institution of Mechanical Engineers, Part
  D: Journal of Automobile Engineering}, vol. 217, no.~1, pp. 41--52, 2003.

\bibitem{yoon2005development}
M.~Yoon, W.~Lee, and M.~Sunwoo, ``Development and implementation of distributed
  hardware-in-the-loop simulator for automotive engine control systems,''
  \emph{International Journal of Automotive Technology}, vol.~6, no.~2, pp.
  107--117, 2005.

\bibitem{kuhlwein2014development}
J.~K{\"u}hlwein, J.~German, and A.~Bandivadekar, ``{Development of test cycle
  conversion factors among worldwide light-duty vehicle CO$_2$ emission
  standards},'' \emph{The International Council on Clean Transportation}, Sept.
  2014.

\bibitem{Lars}
{Eriksson, L.}, ``{Modeling and control of turbocharged SI and DI engines},''
  \emph{Oil \& Gas Science and Technology - Rev. IFP}, vol.~62, no.~4, pp.
  523--538, 2007.

\bibitem{EriNie:2014}
L.~Eriksson and L.~Nielsen, \emph{Modeling and Control of Engines and
  Drivelines}.\hskip 1em plus 0.5em minus 0.4em\relax John Wiley \& Sons, 2014.

\bibitem{ducstegor2006structural}
D.~Düştegör, E.~Frisk, V.~Cocquempot, M.~Krysander, and M.~Staroswiecki,
  ``{Structural analysis of fault isolability in the DAMADICS benchmark},''
  \emph{Control Engineering Practice}, vol.~14, no.~6, pp. 597--608, 2006.

\bibitem{edwards2000sliding}
C.~Edwards, S.~K. Spurgeon, and R.~J. Patton, ``Sliding mode observers for
  fault detection and isolation,'' \emph{Automatica}, vol.~36, no.~4, pp.
  541--553, 2000.

\bibitem{Ng2012}
K.~Y. Ng, C.~P. Tan, and D.~Oetomo, ``Disturbance decoupled fault
  reconstruction using cascaded sliding mode observers,'' \emph{Automatica},
  vol.~48, no.~5, pp. 794--799, 2012.

\bibitem{simani2000diagnosis}
S.~{Simani}, C.~{Fantuzzi}, and S.~{Beghelli}, ``Diagnosis techniques for
  sensor faults of industrial processes,'' \emph{IEEE Transactions on Control
  Systems Technology}, vol.~8, no.~5, pp. 848--855, Sep. 2000.

\bibitem{kobayashi2003application}
T.~Kobayashi and D.~L. Simon, ``{Application of a bank of Kalman filters for
  aircraft engine fault diagnostics},'' in \emph{Proceedings of the ASME Turbo
  Expo 2003, collocated with the 2003 International Joint Power Generation
  Conference. Volume 1: Turbo Expo 2003.}\hskip 1em plus 0.5em minus
  0.4em\relax Atlanta, Georgia, USA: ASME, 2003, pp. 461--470.

\bibitem{darouach1996reduced}
M.~{Darouach}, M.~{Zasadzinski}, and M.~{Hayar}, ``Reduced-order observer
  design for descriptor systems with unknown inputs,'' \emph{IEEE Transactions
  on Automatic Control}, vol.~41, no.~7, pp. 1068--1072, July 1996.

\bibitem{yang1998observer}
H.~Yang and M.~Saif, ``Observer design and fault diagnosis for state-retarded
  dynamical systems,'' \emph{Automatica}, vol.~34, no.~2, pp. 217--227, 1998.

\end{thebibliography}
